\documentclass[onecolumn,aps,floatfix,notitlepage,superscriptaddress,10pt]{revtex4-1}
\usepackage{amssymb,amsmath,amsfonts}
\usepackage{graphics,dcolumn,bm,epic,eepic,float} 
\usepackage{graphicx,subfigure} 
\usepackage{epstopdf} 
\usepackage{bm} 
\usepackage{color} 

\begin{document} 

\title{Identifying the structural discontinuities of human interactions}

\author{Sebastian Grauwin}
\affiliation{Senseable City Lab, Massachusetts Institute of Technology, 77 Massachusetts Avenue, Cambridge, MA 02139, USA}

\author{Michael Szell}
\affiliation{Senseable City Lab, Massachusetts Institute of Technology, 77 Massachusetts Avenue, Cambridge, MA 02139, USA}

\author{Stanislav Sobolevsky}
\affiliation{Senseable City Lab, Massachusetts Institute of Technology, 77 Massachusetts Avenue, Cambridge, MA 02139, USA}

\author{Philipp H\"ovel}
\affiliation{Institut f\"ur Theoretische Physik, Technische Universit\"at Berlin, Hardenbergstra{\ss}e 36, 10623 Berlin, Germany}
\affiliation{Bernstein Center for Computational Neuroscience, Humboldt-Universit{\"a}t zu Berlin, Philippstra{\ss}e 13, 10115 Berlin, Germany}
\affiliation{Center for Complex Network Research, Northeastern University, 110 Forsyth Street, Boston, MA 02115, USA}

\author{Filippo Simini} 
\affiliation{Center for Complex Network Research, Northeastern University, 110 Forsyth Street, Boston, MA 02115, USA}
\affiliation{Department of Engineering Mathematics, University of Bristol, Woodland Road, Bristol, BS8 1UB, United Kingdom}
\affiliation{Institute of Physics, Budapest University of Technology and Economics, Budafoki \'ut 8, Budapest, H-1111, Hungary}

\author{Maarten Vanhoof}
\affiliation{D\'epartement SENSE, Orange Labs, 38 rue du G\'en\'eral Leclerc, 92794 Issy-les-Moulineaux, France}
\author{Zbigniew Smoreda}
\affiliation{D\'epartement SENSE, Orange Labs, 38 rue du G\'en\'eral Leclerc, 92794 Issy-les-Moulineaux, France}

\author{Albert-L\'aszl\'o Barab\'asi}
\affiliation{Center for Complex Network Research, Northeastern University, 110 Forsyth Street, Boston, MA 02115, USA}
\affiliation{Center for Cancer Systems Biology, Dana-Farber Cancer Institute, Boston, Massachusetts 02115, USA}
\affiliation{Department of Medicine, Brigham and Women's Hospital, Harvard Medical School, Boston, Massachusetts 02115, USA}

\author{Carlo Ratti}
\affiliation{Senseable City Lab, Massachusetts Institute of Technology, 77 Massachusetts Avenue, Cambridge, MA 02139, USA}

\date{\today}

\maketitle



{\bf The idea of a hierarchical spatial organization of society lies at the core of seminal theories in human geography that have strongly influenced our understanding of social organization. In the same line, the recent availability of large-scale human mobility and communication data has offered novel quantitative insights hinting at a strong geographical confinement of human interactions within neighboring regions, extending to local levels within countries. However, models of human interaction largely ignore this effect. Here, we analyze several country-wide networks of telephone calls and uncover a systematic decrease of communication induced by borders which we identify as the missing variable in state-of-the-art models. Using this empirical evidence, we propose an alternative modeling framework that naturally stylize the damping effect of borders. We show that this new notion substantially improves the predictive power of widely used interaction models, thus increasing our ability to predict social activities and to plan the development of infrastructures across multiple scales. }\\

Globalization has led us to believe that our world is becoming borderless and deterritorialized. The rise of novel information technologies has even prompted the forecast of the ``death of distance''\cite{cairncross2001ddh}. However, even a most basic organization of society requires categories, compartments and borders to maintain order\cite{newman2006lcs}. Confinement of human interactions to limited spatial areas is the key message of the long-standing hypothesis of Central Place Theory (CPT)\cite{christaller1933dzo,losch1940row} which posits the existence of regular spatial patterns in regional human organization. In short, CPT assumes the existence of a ``hierarchy'' of regions that aims to explain the number, size and locations of human settlements with spatio-economic arguments. Despite its highly simplifying geometric assumptions (Supplementary Information), empirical evidence for CPT's main prerequisite of systematically limited human interactions has been collected in a number of recent studies on massive interaction networks which have indeed observed a substantial impact of political or socio-economic boundaries on human interactions\cite{blondel2010rbm,ratti2010rmg,sobolevsky2013delineating,amini2014impact,szell2012ums,thiemann2010sbs,rinzivillo2012dgb}.  
Typically, if we construct regions by clustering those locations that have strong interactions with each other, we divide countries into contiguous geographical regions with separating boundaries often following surprisingly close existing administrative boundaries. These findings point towards spatial regularities in human organization, prompting us to ask: is there an underlying, rigorously quantifiable, principle behind these patterns? If so, can we exploit this principle to develop better models of human interaction?

\section*{Results}
\subsection*{Quantifying the inhibitory effect of borders on human interaction}
\begin{figure}
\begin{center}
\includegraphics[width=\textwidth]{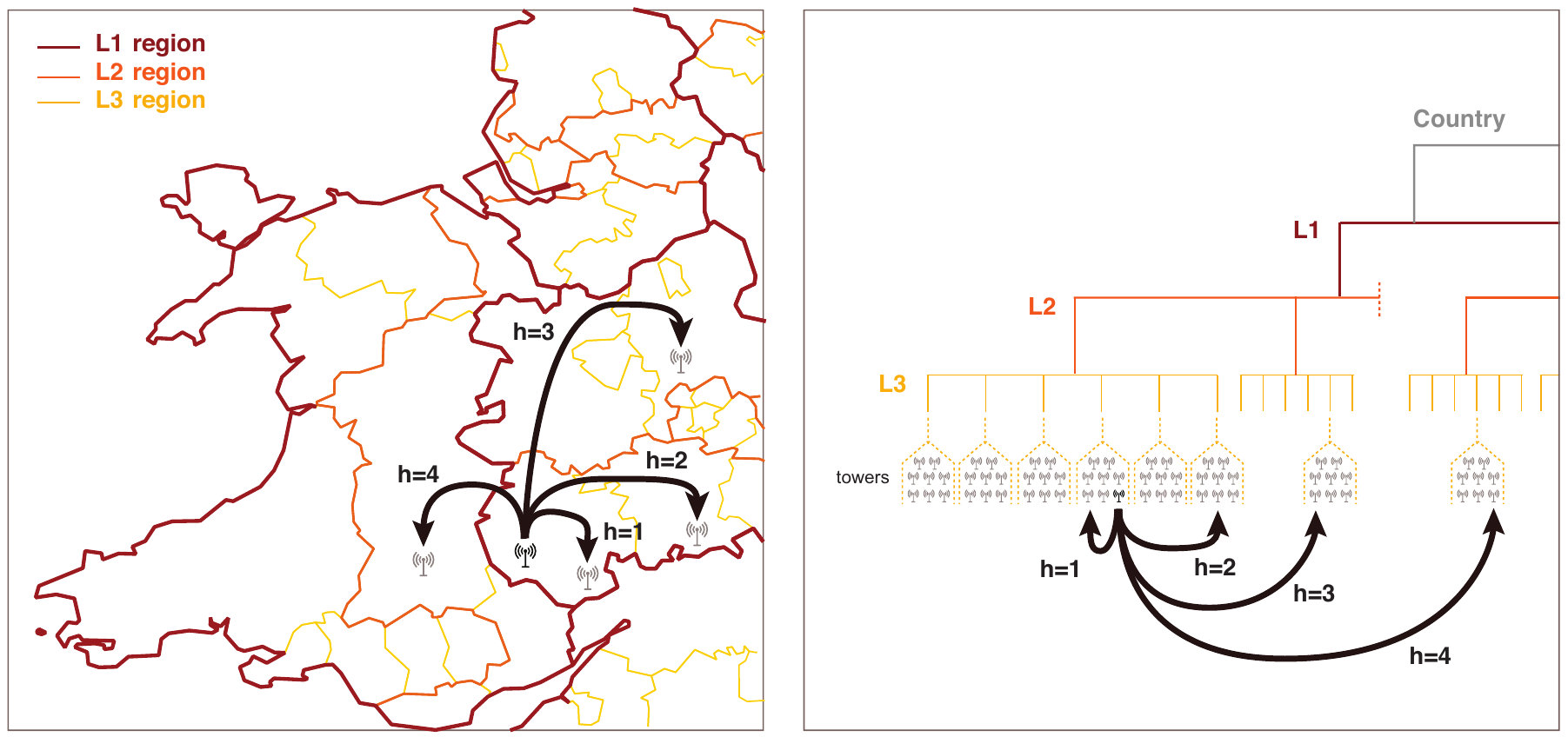}
\end{center}
\caption{{\bf Partitioning of a country based on telephone call networks.}   
Hierarchical distances between two locations are defined through three regional levels - either administrative ones or those found by applying iterative community detection on human interaction networks. Two distinct locations are at a hierarchical distance $h=1$ if they are in the same $L_3$ region, $h=2$ if they are in different $L_3$ regions but in the same $L_2$ region, $h=3$ if they are in different $L_2$ regions but in the same $L_1$ region and $h=4$ if they are in different $L_1$ regions. Note that a higher hierarchical distance does not necessarily correspond to higher geographical distance.
\label{fig1}}
\end{figure}

To quantify the hypothesized effect of hierarchical organization on human interactions, we first define consistent nested regional partitions by recursively applying a community detection algorithm to country-wide phone call networks from the United Kingdom, Portugal, France, Ivory Coast and an anonymous country, Country X (Methods). Partitions resulting from this algorithm reflect the communities defined by underlying social interactions, and, contrary to official administrative boundaries, are independent of country-specific historical or political contexts\cite{sobolevsky2013delineating} (Supplementary Information).
The resulting partition consists of three levels, $L_1$, $L_2$, and $L_3$, that have a natural interpretation: the whole country is divided into $L_1$-level regions (regional scale), which are divided into $L_2$-level regions (county scale) which in turn split into $L_3$-level regions (city scale) composed of several ``elementary'' locations (cell phone tower or exchange area), Fig.~\ref{fig1}. The number of levels is not imposed, but for all countries the process naturally stops subdividing regions at the city scale. 
To our surprise, we find that, although no spatial constraints are applied, communities consist of contiguous locations at all levels (an observation that had previously been reported just for $L_1$-level regions\cite{ratti2010rmg,sobolevsky2013delineating}), and are strikingly similar to administrative regions as highlighted by comparison with random partitions (Supplementary Table 1).

Several insights that we first derive from these hierarchical partitions of empirical networks are in line with CPT. The $L_3$ regions have typical spatial extension of a town with its neighborhood\cite{haggett1977locational} (between $15\,$km to $23\,$km, depending on the country); similarly, $L_2$ and $L_1$ conform to the scale of districts and regions (Fig.~\ref{fig2}a). The distribution $P(n)$ of the number $n$ of $L_3$ communities inside a $L_{2}$ community is strongly peaked (around $6$) providing quantitative confirmation to the main hypothesis of regular spatial organization of CPT (Supplementary Information). Figure~\ref{fig2}f shows that distribution, as well as the distribution of the number of $L_2$ communities within an $L_{1}$ community ($\#L2/\#L1$), for the UK. We observe similar peaks in all other countries ( Figs.~S1--S4). 
Following the idea that borders inhibit human interaction, we introduce the notion of hierarchical distance to characterize their impact on communication flows (Fig.~\ref{fig1}). Two locations $i$ and $j$ are at a hierarchical distance $h_{ij}=1$ if they are in the same $L_3$ region, $h_{ij}=2$ if they are in different $L_3$ regions, but in the same $L_2$ region, $h_{ij}=3$ if they are in different $L_2$ regions, but in the same $L_1$ region and $h_{ij}=4$ if they are in different $L_1$ regions. In other words, the hierarchical distance corresponds to the number of different types of borders separating two locations. This metric only contains limited information about the spatial structure of the regions and is only partly correlated with geographic distance: two locations that are close in terms of geographic distance can still be situated in two distinct $L_1$ regions and hence far from each other in terms of hierarchical distance. 
Thus, the hierarchical distance is not a mere discretization of geographical distance, but encodes a qualitatively different, socio-economic notion of distance. To understand the impact of borders on human interaction on each hierarchical level, we define and measure the following damping parameters
\begin{equation}
q_i^{(h)} =  \frac{T^{(h+1)}_{i}}{W^{(h+1)}_{i}} \frac{W^{(h)}_i}{T^{(h)}_i}, \qquad h=1,2,3,
\end{equation}
where $T^{(h)}_{i} =  \sum_{j: h_{ij}=h} T_{ij}$ is the total duration of calls from location $i$ to all locations at hierarchical distance $h$. Defining the weight of node $i$, $w_i=\sum_j T_{ij} $, as the total duration of calls originating from node $i$ (including self-loops), $W^{(h)}_i = \sum_{j: h_{ij}=h} w_j$ is the total duration of calls originating from locations at a hierarchical distance $h$ from location $i$. The ratio $T^{(h)}_{i} / W^{(h)}_{i}$ measures the relative strength of communication between location $i$ and the locations at a hierarchical distance $h$ from it. 
In particular, this ratio corresponds to the amount of communication sent to all locations at hierarchical distance $h$ per unit of communication produced there.
The damping value $q_i^{(h)}$ hence measures the relative importance of locations at hierarchical distance $h+1$ compared to those at hierarchical distance $h$ from $i$.
For example, $q_i^{(h)} = 1$ for all $h$ means that communication from $i$ is independent of the hierarchical distance, because there is no damping in the amount of communication sent per unit of communication produced as the hierarchical distance increases.
Figure \ref{fig2}k shows the distributions of the damping values in the UK, all well peaked around a strikingly similar mean value which does not substantially change with the hierarchy level ($h=1,2,3$), Table \ref{tab:damping}. Similar observations are made for all studied countries (Figs.~S1--S4). This finding - signing a structural discontinuity of human interactions - and its consequences on modeling, see below, is our main discovery. It means that the damping effect of a boundary is approximately the same irrespective of the level $h$ and origin location $i$, i.e. $q_i^{(h)}\simeq q$.
If the probability for two people who live in the same $L_3$ region to communicate is $p_0$, it will be $qp_0$  for people living in different $L_3$ regions but in the same $L_2$ region, $q^2p_0$ if they live in the same $L_1$ but different $L_2$ region and $q^3p_0$ if they live in different $L_1$ regions, Fig. \ref{fig4}b.
 
\subsection*{Why and how standard models fail}
\begin{figure}
\begin{center}
\includegraphics{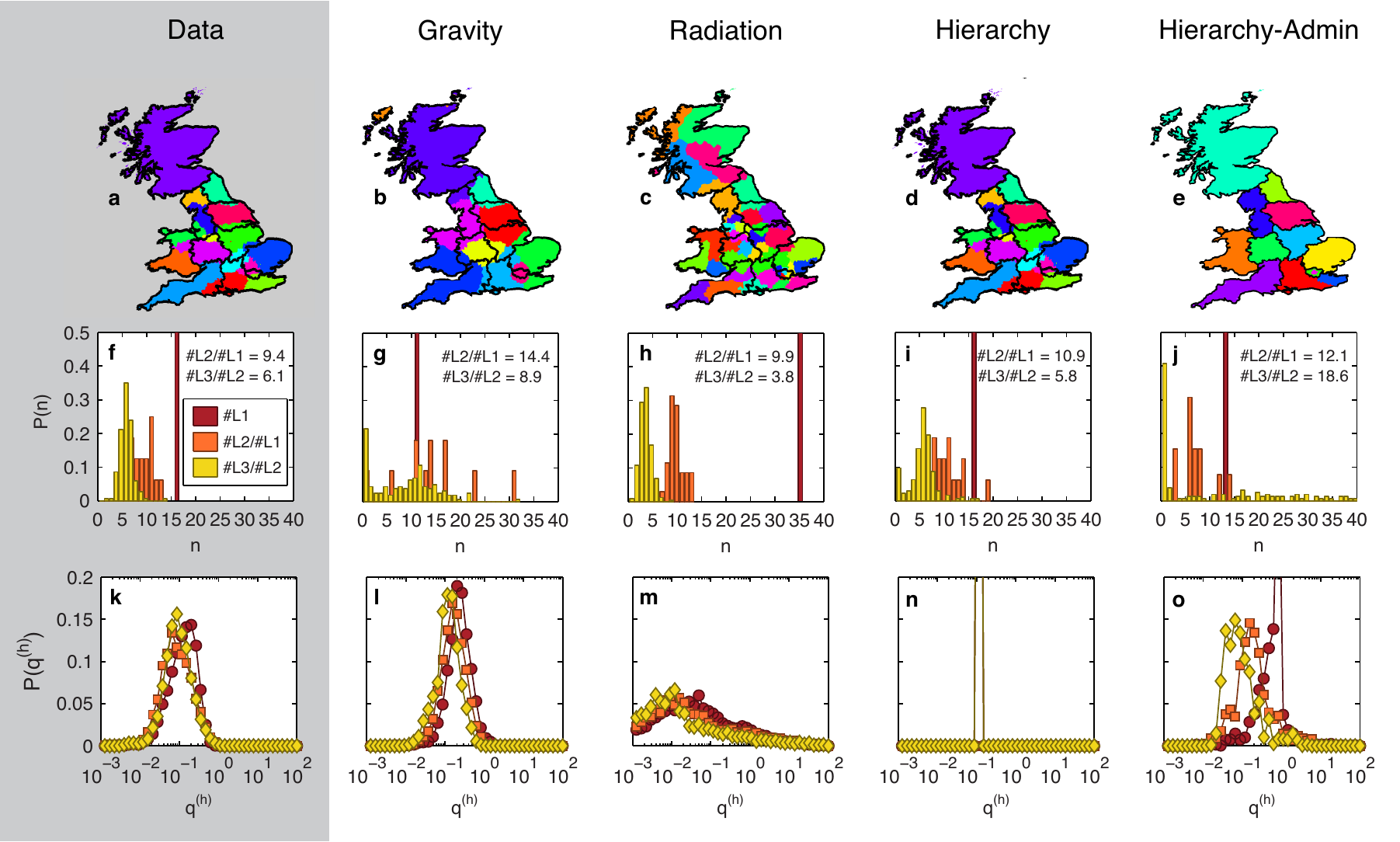}
\end{center} 
\caption{{\bf Hierarchical properties of spatial organization from human interactions.} {\bf a--e}, Maps of $L_1$ communities in telephone call networks detected from data and from various interaction models. Black lines correspond to the administrative partitioning of the 11 NUTS1 regions of UK, colored areas to regions detected by a community detection algorithm applied to ({\bf a}) the data, and to the ({\bf b}) gravity, ({\bf c}) radiation, ({\bf d}) hierarchy, and ({\bf e}) administrative models. All detected regions are cohesive although some of the distinct colors used may appear similar.
{\bf f--j}, Probability distribution of number of subregions by region found in ({\bf f}) the actual network and ({\bf g--j}) in each model. The gravity model ({\bf g}) underestimates the number of $L_1$ communities but overestimates the numbers of subregions within regions. The radiation model ({\bf h}) strongly overestimates the number of $L_1$ communities. The hierarchy model ({\bf i}) correctly determines the distributions of sub-communities per community.
{\bf k--o}, Probability distributions of damping values $q^{(h)}$. The hierarchy model ({\bf n}) assumes a constant damping value for all levels.
\label{fig2}} 
\end{figure}

\begin{figure}
\begin{center}
\includegraphics{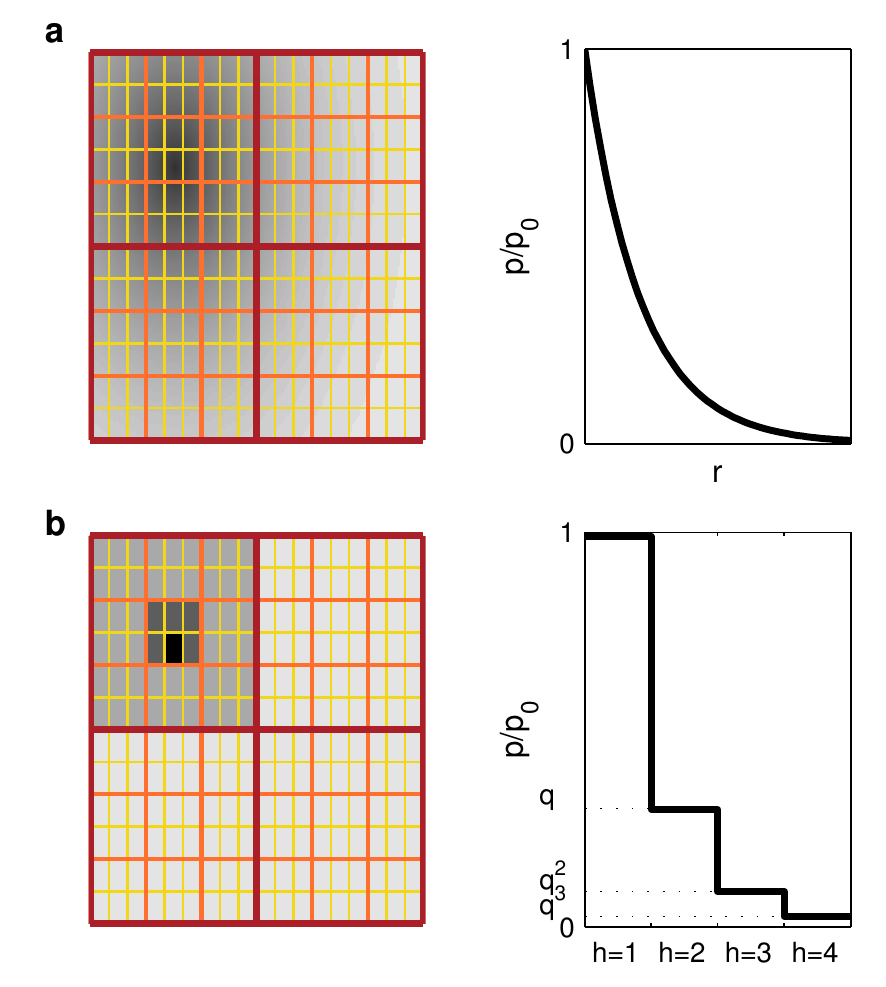}
\end{center} 
\caption{{\bf Schematic representation of the structural discontinuity effect.} {\bf a,} In the classic gravity model, the probability $p$ that two people communicate is a continuous (e.g. exponential) function of the distance between them. {\bf b,} In our hierarchy model, that probability is a discontinuous function induced by the assumption of a constant damping value $q$ independent of the point of origin and the hierarchy level $h$. In both cases, the left panel shows in grayscale the probability of communication from a given point in space in a schematic country, partitioned in three regional levels with the same color coding as Fig \ref{fig1}. The link between the borders and the deterrence function is clearly apparent in the second case. \label{fig4}} 
\end{figure}

Using a standard quality of fit statistic (Methods) and comparing distributions of high level per low level regions, we tested to which extent the most widely used models, namely gravity\cite{zipf1946p} and radiation\cite{simini2012umm}, commit a systematic bias by failing to account for the observed boundary effects. To this end, we compute the communication networks predicted by these models as well as the corresponding partitions resulting from the community detection algorithm (Methods). As previously demonstrated elsewhere\cite{simini2012umm}, the gravity model strongly underestimates and fails to predict high-range flows, i.e.~flows between locations where the number of calls is high (Fig.~\ref{fig3}a andFigs.~S5a to S8a). This certainly explains why the gravity model generates less and larger $L_1$ regions whose subdivisions do not follow the narrow distributions observed in the data (Fig.~\ref{fig2}b and~\ref{fig2}g). The damping value predicted by the gravity model is otherwise well peaked, although its average values vary from one $h$ level to another (Table~\ref{tab:damping}). In contrast, the radiation model overestimates long-range flows (Fig.~\ref{fig3}b), resulting in more and smaller $L_1$ regions (Fig.~\ref{fig2}c). Therefore, the distribution of $L_3$ within $L_2$ regions is, although well-peaked, shifted to the left (Fig.~\ref{fig2}h). The distribution of damping values in the radiation model is moreover strongly spread out (Fig.~\ref{fig2}m and Table~\ref{tab:damping}), and does not reproduce the existence of a single typical damping parameter. Similar systematic biases of gravity and radiation models become evident if we measure the probability $P_{\mathrm{dist}}(d)$ of a call between locations at distance $d$ (Fig.~S9).

\subsection*{Accounting for strong border effects with the Hierarchy model}
The two most commonly used models thus fail to reproduce the boundary effect.
By design, a model taking into account the observed hierarchical structures by assuming a constant damping value $q$, would overcome this issue (Fig. \ref{fig4}b). Consider the minimal model in the stylized form $T_{ij} \propto N_{ij} q^{h_{ij}}$, where $N_{ij}$ represents the potential pairs of contacts between two distinct locations $i$ and $j$ and $q^{h_{ij}}$ the probability for two people from these locations to communicate. This model would implement highly discretized hierarchical distances instead of considering a continuum of geographical distances. Similarly to the gravity model, $N_{ij}$ can be taken as proportional to the weight $w_i$ and $w_j$ of both origin and destination locations. We therefore propose a simple \emph{hierarchy model} that predicts an interaction strength as 
\begin{equation}
\label{eq:hiermodel}
T_{ij}^{Hier} = C_i w_i w_j q^{h_{ij}},
\end{equation}
where $0 < q < 1$ is a parameter to be determined and $C_i$ are local normalization factors ensuring $w_i^{Hier} = w_i$. This normalization also ensures that the damping values are constants, $q_i^{Hier} = q$ (Supplementary Information).
The best-fit values of $q$ are very close to the observed values (Table \ref{tab:damping} and Table S2) and robust to small variation (Fig.~S10). These values slightly depend on the country, varying between $0.10$  and $0.25$, reflecting differences in the structural properties of the studied networks. The hierarchy model reproduces almost perfectly the nested structure of regions (Figs.~\ref{fig2}d and \ref{fig2}i). To our surprise, the hierarchy model also outperforms the state-of-the-art models in terms of goodness of fit measures (Table \ref{tab:benchmark} and Extended Table 3). In particular, it estimates high-range flows with a greater accuracy than the radiation or gravity models, as can be seen on the top right corners of Fig.~\ref{fig3}c and Figs.~S5c to S8c, where the markers are typically closer to the equality line than in state-of-the-art models.

\section*{Discussion} 
\begin{figure}
\begin{center}
\includegraphics{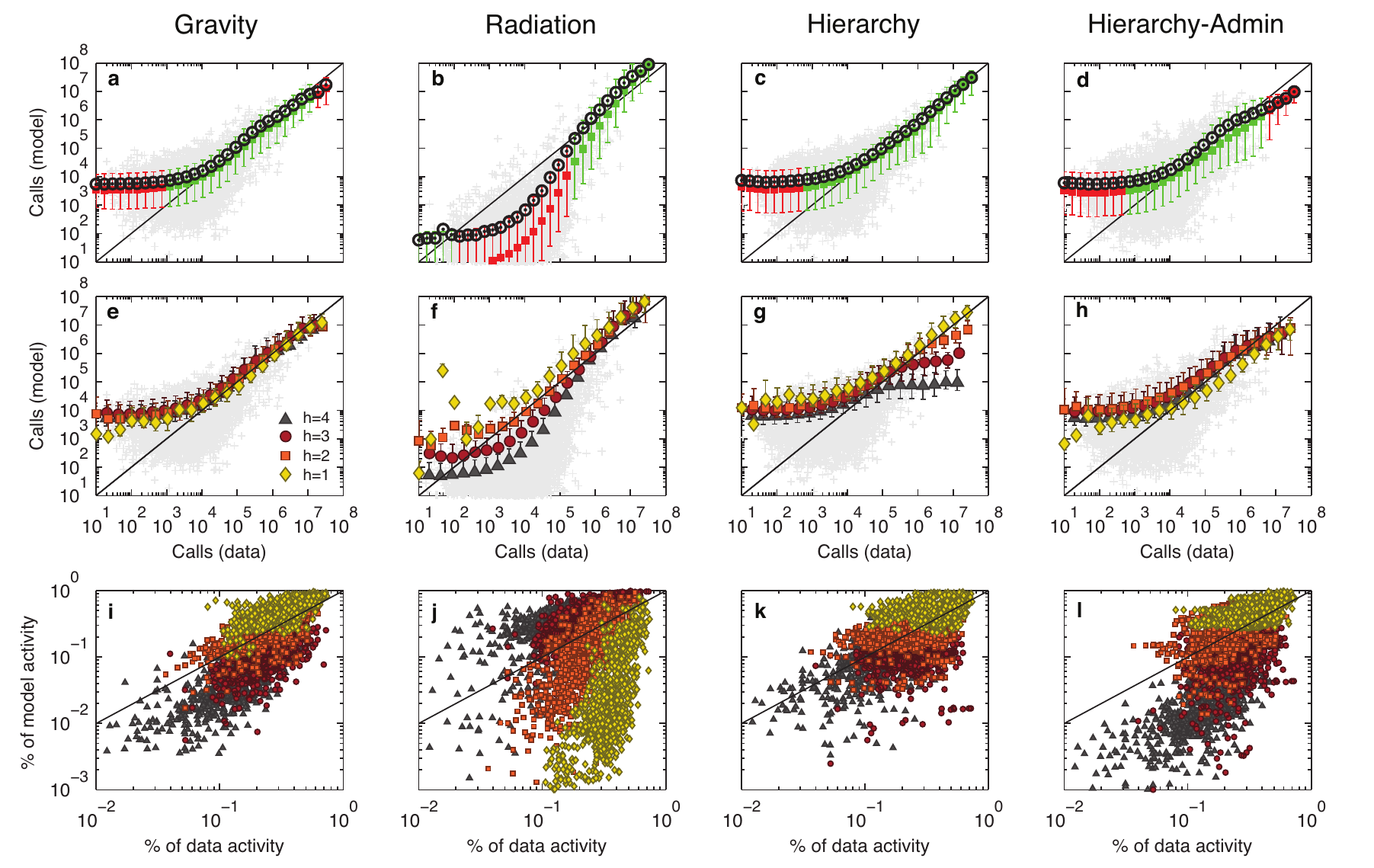}
\end{center} 
\caption{{\bf Comparison of model predictions.} {\bf a--d}, Comparison of the actual total communication to the predicted communication for each pair of distinct locations, for the ({\bf a}) gravity, ({\bf b}) radiation, ({\bf c}) hierarchy, and ({\bf d}) administrative models. Gray markers are scatter plots for each pair of locations. A box is colored green if the equality line $y=x$ lies between the 9th and 91th percentiles in that bin and is red otherwise. Red boxes hence emphasize significant biases of the models. Black circles correspond to the average total communication of the pairs of locations in that bin.
{\bf e--h}, Goodness of prediction with respect to the hierarchical distance $h$, for the ({\bf e}) gravity, ({\bf f}) radiation, ({\bf g}) hierarchy, and ({\bf h}) administrative models. Gray markers are scatter plots for each pair of locations. Error bars show the corresponding 9th and 91th percentiles of total communication values. {\bf i--l}, For each L3 community, comparison of the fractions of activity of model versus data between that L3 community and L3 communities at different hierarchical distances, for the ({\bf i}) gravity, ({\bf j}) radiation, ({\bf k}) hierarchy and ({\bf l}) administrative models. 
\label{fig3}} 
\end{figure}

Focusing on flows between locations at specific hierarchical distance, the goodness of prediction of the different models is informative to understand why the hierarchy model outperforms the others. 
The radiation model overestimates the flows at $h=1,2$, the corresponding markers being above the equality line in Fig.~\ref{fig3}f and Figs.~S5f to S8f resulting in an overall overestimation quantified by values of $R_{h=1,2}$ (Methods) greater than 1 (Table~\ref{tab:ratios}), and underestimates those at $h=3,4$. On the contrary, the gravity model underestimates the flows at $h=1,2$ and overestimates those at $h=3,4$ (Table~\ref{tab:ratios}, Fig.~\ref{fig3}e, Figs. S5e to S8e).
The hierarchy model produces more balanced predictions ($R_h$ closer to 1) and thus outperforms existing models.

The hierarchy model requires the knowledge of the communication flows in order to determine the three hierarchical levels each location belongs to. However, it can also be applied in the absence of communication data, using the administrative boundaries and a general damping value $q=0.2$ which matches robustly all countries (Extended Data Fig.~10). The resulting \emph{hierarchy-admin} model based on this administrative partition, is parameter-free and yet it provides similar or sometimes better estimates than the gravity model in terms of communication flow (Fig.~\ref{fig3}k, Figs.~S5k to S8k: see in particular the case of high-range flow in Portugal and Ivory Coast) or benchmark measures (Table \ref{tab:benchmark} and Extended Data Table 3). We also tested different constraint conditions and deterrence functions $f$ in the hierarchy model $T_{ij}^{Hier} = C_i w_i^\alpha w_j^\beta f(h_{ij})$. We compared them to multiple variations of the gravity and radiation models which are widely outperformed by hierarchy models (Supplementary Information).

In summary, we first defined communication flows-induced boundaries by applying standard community detection methods on large-scale human interaction networks and found that these networks have a nested structure reflecting historic, socio-political borders which can be related to the structure predicted by CPT. 
We introduced the notion of damping parameter, representing the normalized ratio of interactions between locations at different hierarchical distances, to quantify the inhibiting effect of boundaries. Surprisingly, the distributions of damping parameters are well-peaked and largely independent of the hierarchical level, revealing a structural discontinuity effect in each studied country. We further showed that current models of human interaction, based only on population and/or geographical distance, cannot correctly reproduce the characteristic hierarchical structure of interaction networks. We proposed a simple model based on the discrete hierarchical distance that outperforms the state-of-the-art models of human interaction in a number of different countries, demonstrating its general applicability and emphasizing the impact of the borders on human interactions. The development of more sophisticated models combining both geographic and socio-political information will further boost our ability to understand and reproduce the structure of social systems.


\section*{Methods}
\subsection*{Telephone call data} 
We consider several country-wide data sets of telephone calls, including the four European countries of the UK, France, Portugal and an anonymized Country X, and Ivory Coast.  All data sets comprise mobile phone data with the exception of landline calls in the UK. Data was provided by single phone providers with possibly heterogeneous coverage over the respective countries - we have no information on local market shares and on resulting possible inhomogeneities in spatial coverage. Specific details of the different datasets are provided in Extended Data Table 1, all of them gathering millions of users making billions of calls during time frames ranging from 1 to 15 months. 
We construct interaction networks between different locations of a country based on the aggregated duration of calls having origin in the first and destination in the second location. 
This process generates weighted directed networks in which loop edges from locations to themselves are also considered, and where the link weight $T_{ij}$ between a location $i$ and location $j$ is defined as the total duration (or, in case of Country X, total number) of calls from location $i$ to location $j$.  The nodes of the network are the locations, corresponding to exchange areas or cell towers areas as reported in Extended Data Table 2. In all datasets, the users are attached to the actual locations where the calls occur (i.e. not necessarily their residential locations).

\subsection*{Network partitioning} 
A recently developed algorithm for community detection, referred to as ``Combo''\cite{ratti2010rmg,sobolevsky2013delineating,amini2014impact} is applied to the extracted communication networks to detect communities of highly connected locations. The method follows a standard modularity optimization approach\cite{newman2006mac, fortunato2010report}, scoring the edges of the networks according to their relative strength compared to a null-model based on the weight of the nodes they connect and aiming at maximizing the cumulative score inside the communities. 
Given a partition of the nodes in a set of clusters ${c_i}$, the modularity score $Q$ is given by
\begin{equation}
Q = \frac{1}{W}\sum_{ij} \left[T_{ij}- \frac{w_iw_j}{W} \right] \delta(c_{i}, c_{j}),
\end{equation}
where $T_{ij}$ is the weight of the link between node $i$ and node $j$, $w_i=\sum_j T_{ij}$ is the weight of node $i$ and $W=\sum_i w_i / 2$ is the total weight of the network.
While the outcome of partitioning networks is in general not qualitatively dependent on the particular algorithm used, the Combo algorithm has the ability to consistently provide the best results in terms of modularity score compared to other algorithms\cite{sobolevsky2013general}.
The modularity optimization approach yields communities whose size and properties are only based on the informations of the links' weights. See\cite{fortunato2007resolution} for a more explicit interpretation of the modularity, its properties and limits.

Applying the Combo algorithm yields a first partition of the network into communities, further referred to as ``level 1'' or ``$L_1$'' partition. To obtain the substructure of these communities, we iteratively apply the Combo algorithm on each $L_1$ community, thus creating a``level 2'' or ``$L_2$'' community partition, and and then again on each $L_2$ community, thus creating a ``level 3'' or ``$L_3$'' community partition. We find that most of the $L_1$ and $L_2$ communities display a clear substructure with high values of internal modularity scores, typically around 0.4 and 0.7 (Supplementary~Table~I). The resulting communities consists in geographically cohesive regions, which can seem surprising since the algorithm uses only the networks topology and no geographical information, such as the distance between the nodes (Supplementary). This cohesiveness is also linked to the spatial scale of the studied network: we would not expect any contiguous communities if that analysis was done at a city scale, where the movements and communications of individuals are more evenly distributed in space.

\subsection*{Interactions models and goodness measures} 
The radiation model is a parameter-free model recently introduced in the context of migration patterns\cite{simini2012umm}. Given the geographic distance $d_{ij}$ between two locations $i$ and $j$, the model predicts that the flow of individuals moves $T_{ij}$ between these two distinct locations will depend on the population at the origin, the population at the destination and on the population $s_{ij}$ within the circle of radius $d_{ij}$ centered on the origin location $i$. Applied to our case (using the total communication $w_i$ at location $i$ as a proxy for its population), the radiation model is written as 
\begin{equation} 
T_{ij}^{Radiation} = C_i \frac{w_i w_j}{(w_i + s_{ij})(w_i + w_j+ s_{ij})},
\label{EQ:RM} 
\end{equation} 
where $s_{ij}=\sum_{k,\, 0<d_{ik}<d_{ij}}w_k$ is the total amount of communication originating from locations at a distance shorter than $d_{ij}$ from location $i$ and $C_i$ is a normalization factor ensuring that the predicted total activity of each node is the same than the actual one, i.e. $\sum_{j\neq i} T_{ij}^{Radiation}= \sum_{j\neq i} T_{ij}$. The model is otherwise parameter-free.  
 
The gravity model is one of the oldest models describing human mobility and interaction, formulated in analogy to Newton's law of gravity. The classical form predicts that the interaction strength between two distinct locations varies with the distance between them according to a power law:
\begin{eqnarray}
T_{ij}^{Gravity} &=&  C w_i^{\alpha} w_j^{\beta} d_{ij}^{\gamma}, 	\label{EQ:GplM}
\end{eqnarray} 
where $C$ is a global normalization constant ensuring that $\sum_{i,\, j\neq i} T_{ij}^{Gravity}= \sum_{i,\,j\neq i} T_{ij}$ and $\alpha$, $\beta$, and $\gamma$ are parameters to fit. 

We also computed the generalized version of the radiation model proposed in\cite{simini2013human}, as well as different versions of the gravity and hierarchy models, comparing the results obtained using a power law or exponential deterrence function (Supplementary Information). All parameters in these models were estimated through a regression analysis minimizing the deviance $E$ \cite{nelder1972generalized}, a measure based on the log-likelihood of model compared to a sturated model that can be interpreted as a generalization of the residual sum of squares $R^2$.

We also quantify the fits between communication networks and models through different benchmark measures, namely the Dice distance $D$, the Sorensen distance $S$, and the cosine distance $C$ defined by: 
\begin{eqnarray} 
D(A,M) &=& \frac{\sum_{ij}\left( M_{ij} - A_{ij} \right)^2}{\sum_{ij}M_{ij}^2+\sum_{ij}A_{ij}^2}   \\
S(A,M)&=&\frac{\sum_{ij}| M_{ij} - A_{ij} |}{\sum_{ij} \left( M_{ij} + A_{ij}\right)}  \\ 
C(A,M) &=& 1- \frac{\sum_{ij} M_{ij}A_{ij}}{\sqrt{\sum_{ij} M_{ij}^2} \sqrt{\sum_{ij}A_{ij}^2}}.
\label{EQ:fitDistance} 
\end{eqnarray}  
These three benchmark measures cover most families of distance measures\cite{cha2007comprehensive}, which allows us to ensure that our findings are stable with respect to the distance measure used. They all vary between 0 and 1 and the lower they are, the more similar the model is to the original data. 

Finally, we also computed the correlation $corr$ between each model and the data defined by
\begin{equation} 
corr(A,M) = \frac{\sum_{ij} (M_{ij}-\langle M_{ij} \rangle)(A_{ij}-\langle A_{ij}\rangle)} {\sqrt{\sum_{ij}  (M_{ij}-\langle M_{ij} \rangle)^2} \sqrt{\sum_{ij} (A_{ij}-\langle A_{ij} \rangle)^2}} 
\label{EQ:fitCorr} 
\end{equation}  
which is a measure of similarity varying between -1 and 1 (the closer to 1, the higher the similarity).

\subsection*{Over- and underestimation measure} 
In order to determine whether a given subset of links are over- or underestimated by the models, we define for
any given set $E$ of links, the following ratio: 
\begin{equation} 
R_E(A,M) = \frac{\sum_{ij \in E} M_{ij} }{ \sum_{ij \in E} A_{ij}}.
\end{equation}  
where we use the notation $A$ for the data and $M$ for the model. Values of $R_E$ larger (resp. smaller) than 1 hence correspond to an overestimation (resp. underestimation) of the model. The measure $R_E$ provides an aggregated knowledge dominated by link weights.


\clearpage
\section*{Acknowledgments}
\begin{itemize}
\item P.H. acknowledges support of the German Academic Exchange Service (DAAD) via a postdoctoral fellowship. F.S. acknowledges support of the European FET-Open Project DATASIM (FP7-ICT-270833). S.G., M.S., S.S., and C.R. thank the National Science Foundation, the AT\&T Foundation, the MIT SMART program, the Center for Complex Engineering Systems at KACST and MIT, Volkswagen Electronics Research Lab, BBVA, The Coca Cola Company, Ericsson, Expo 2015, Ferrovial, the Regional Municipality of Wood Buffalo and all the members of the MIT Senseable City Lab Consortium for supporting the research. The authors also thank Dashun Wang, Markus Schl\"apfer, Oleguer Sagarra and Santi Phithakkitnukoon for valuable feedbacks.
\item All authors designed the research, contributed to the model development and edited the manuscript; Z.S. provided the France dataset and A.-L.B. Country X dataset; C.R. proposed a general idea of the project; S.S. provided the concept of a model based on hierarchical distance and itÕs initial validation; F.S. proposed the concept of damping parameter; S.G., P.H. and M.V. processed and cleaned data; S.G. and S.S. designed the algorithms; S.G., M.S. and M.V. implemented the algorithms; S.G., M.S., F.S. and P.H. analyzed the results; S.G. and M.S. were the lead writers of the manuscript.
\item The authors declare that they have no competing financial interests. Correspondence and requests for materials should be addressed to S.G.~(email: sebgrauwin@gmail.com).
\end{itemize}

\clearpage

\begin{table}
\begin{center} 
\caption{ {\bf Values of the damping value $q$} for the actual and modeled networks in the UK. \label{tab:damping}}
\begin{tabular}{l | ccc } 
Data set / Network & $\langle q^{(1)} \rangle$ &  $\langle q^{(2)} \rangle$ &$ \langle q^{(3)} \rangle$ \\ \hline  
Data & 0.180$\pm$0.002 $\quad$& 0.143$\pm$0.002 $\quad$& 0.144$\pm$0.002\\
Gravity & 0.331$\pm$0.005 $\quad$& 0.234$\pm$0.003$\quad$ & 0.167$\pm$0.002\\
Radiation & 8.180$\pm$6.039 $\quad$& 6.156$\pm$3.922 $\quad$& 3.753$\pm$1.687\\
Hierarchy &  0.139$\pm$0.000 $\quad$ & 0.139$\pm$0.000 $\quad$ & 0.139$\pm$0.000\\
Hierarchy-Admin &  0.2$\pm$0.0 $\quad$ & 0.2$\pm$0.0 $\quad$ & 0.2$\pm$0.0
\end{tabular}
\end{center}
\end{table}

\begin{table}
\begin{center}
\caption{ {\bf Benchmark measures quantifying the goodness of fit in the UK.} \label{tab:benchmark}
The Dice (D), Sorensen (S), Cosine (C) and deviance (E) are four different measures of the distance between the actual and modeled networks. The correlation $corr$ measures a similarity between a model and the data. The parameters of the gravity and hierarchy models were chosen to minimize the value of E.}
\begin{tabular}{l | c | ccc | c | l}
    Model & E$\times10^{-12}$ & D & S & C & corr & parameters\\ 
    \hline
    Gravity  & 0.494   &  0.456 & 0.448 & 0.456 & 0.543 & $\alpha=0.65$, $\beta =0.65$, $\gamma=-1.46$\\     
    Radiation &  1.622 &  0.624 & 0.632 & 0.344 &  0.656  &  \\     
    Hierarchy & 0.464 & 0.233 & 0.437 & 0.231 & 0.768 & $q=0.139$\\
    Hierarchy-Admin & 0.679 & 0.503 & 0.527 & 0.458 & 0.540 & $q=0.2$ (imposed)\\
\end{tabular}
\end{center}
\end{table}

\begin{table}
\caption{\bf{Over- / under-estimation measures of link at specific hierarchical distance in the UK.} \label{tab:ratios}}
\begin{center}
\begin{tabular}{l | r r r r | }
    Model  & $R_{h=1}$ & $R_{h=2}$ & $R_{h=3}$ & $R_{h=4}$ \\ \hline
    Gravity  &  0.54 & 0.73 & 1.15 &  1.33\\
    Radiation  & 2.39 & 1.47 & 0.67 & 0.16  \\
    Hierarchy & 1.10 & 0.73 & 0.90 & 1.18\\
    Hierarchy-Admin & 0.25 &0.73 & 1.43 & 1.30
\end{tabular}
\end{center}
\end{table}

\clearpage

\end{document}


\title{Identifying the structural discontinuities of human interactions \\ - \\ Supplementary Informations}

\author{Sebastian Grauwin}
\affiliation{Senseable City Lab, Massachusetts Institute of Technology, 77 Massachusetts Avenue, Cambridge, MA 02139, USA}

\author{Michael Szell}
\affiliation{Senseable City Lab, Massachusetts Institute of Technology, 77 Massachusetts Avenue, Cambridge, MA 02139, USA}

\author{Stanislav Sobolevsky}
\affiliation{Senseable City Lab, Massachusetts Institute of Technology, 77 Massachusetts Avenue, Cambridge, MA 02139, USA}

\author{Philipp H\"ovel}
\affiliation{Institut f\"ur Theoretische Physik, Technische Universit\"at Berlin, Hardenbergstra{\ss}e 36, 10623 Berlin, Germany}
\affiliation{Bernstein Center for Computational Neuroscience, Humboldt-Universit{\"a}t zu Berlin, Philippstra{\ss}e 13, 10115 Berlin, Germany}
\affiliation{Center for Complex Network Research, Northeastern University, 110 Forsyth Street, Boston, MA 02115, USA}

\author{Filippo Simini} 
\affiliation{Center for Complex Network Research, Northeastern University, 110 Forsyth Street, Boston, MA 02115, USA}
\affiliation{Department of Engineering Mathematics, University of Bristol, Woodland Road, Bristol, BS8 1UB, United Kingdom}
\affiliation{Institute of Physics, Budapest University of Technology and Economics, Budafoki \'ut 8, Budapest, H-1111, Hungary}

\author{Maarten Vanhoof}
\affiliation{D\'epartement SENSE, Orange Labs, 38 rue du G\'en\'eral Leclerc, 92794 Issy-les-Moulineaux, France}
\author{Zbigniew Smoreda}
\affiliation{D\'epartement SENSE, Orange Labs, 38 rue du G\'en\'eral Leclerc, 92794 Issy-les-Moulineaux, France}

\author{Albert-L\'aszl\'o Barab\'asi}
\affiliation{Center for Complex Network Research, Northeastern University, 110 Forsyth Street, Boston, MA 02115, USA}
\affiliation{Center for Cancer Systems Biology, Dana-Farber Cancer Institute, Boston, Massachusetts 02115, USA}
\affiliation{Department of Medicine, Brigham and Women's Hospital, Harvard Medical School, Boston, Massachusetts 02115, USA}

\author{Carlo Ratti}
\affiliation{Senseable City Lab, Massachusetts Institute of Technology, 77 Massachusetts Avenue, Cambridge, MA 02139, USA}

\date{\today}

\maketitle

\section*{Supplementary Tables}

\begin{table}[h!]
\begin{center}
\caption{{\bf (Table S1)}  {\bf Properties of the data sets.} {\footnotesize Country-wide telephone data sets are provided by single telephone operators, covering different time frames, with different numbers of phones, calls, total call durations and on various spatial resolutions. The abbreviations bn. and m. stand for billion and million, respectively. Resolution numbers are given as approximate values. These locations constitute the nodes of the corresponding telephone call networks, while the sum of durations of calls between locations span their weighted links. The last columns report the percentage of non-zero links between pairs of nodes in the extracted network and whether to not that network is directed. The durations of calls are unknown in the case of Country X. All datasets corresponds to mobile phone network except for UK, where the dataset corresponds to a landline network.} \label{tab:datasets} }
\vspace{2mm}
\begin{tabular}{c | c | c | c | c | c | c | c | c }
    Data set  & $\,\,$ Calls $\,\,$ &  $\,\,$ Duration (s)  $\,\,$ &  $\,\,$Phones $\,\,$ &  $\,\,$Time $\,\,$ &  $\,\,$Spatial resolution $\,\,$ &  $\,\,$$\rho_{links}$ $\,\,$ &  $\,\,$Directed $\,\,$ \\ 
    \hline
    France  & 218 m. & 47 bn. & 17.6 m. & 1 month & 8,800 areas & $11.6\%$ & yes\\
    UK  & 7.6 bn. & 452 bn. & 47 m. & 1 month & 4,800 areas &  $37.6\%$ & no\\
    Portugal  & 440 m. & 56 bn. & 1.6 m. & 15 months & 2,200 cell towers &  $83.1\%$ & yes \\
    Country X  & 1.1 bn. & - & 6.9 m. & 12 months & 9,400 cell towers & $28.0\%$ & yes\\
    Ivory Coast  & 62 m. &7 bn. & 5m. & 6 months & 1,250 cell towers & $84.2\%$ & no\\
\end{tabular}
\end{center}
\end{table}

\begin{table}
\begin{center} 
\caption{ {\bf (Table S2)} {\bf Values of the damping parameter $q$} for the actual and modeled networks in France, Portugal, Country X and Ivory Coast. \label{tab:dampingmod}}
\begin{tabular}{l | ccc } 
Data set / Network & $\langle q^{(1)} \rangle$ &  $\langle q^{(2)} \rangle$ &$ \langle q^{(3)} \rangle$ \\ \hline
UK / Data & 0.180$\pm$0.002 $\quad$& 0.143$\pm$0.002 $\quad$& 0.144$\pm$0.002\\
UK / Gravity & 0.331$\pm$0.005 $\quad$& 0.234$\pm$0.003$\quad$ & 0.167$\pm$0.002\\
UK / Radiation & 8.180$\pm$6.039 $\quad$& 6.156$\pm$3.922 $\quad$& 3.753$\pm$1.687\\
UK / Hierarchy &  0.139$\pm$0.000 $\quad$ & 0.139$\pm$0.000 $\quad$ & 0.139$\pm$0.000\\ 
UK / Hierarchy-Admin &  0.2$\pm$0.0 $\quad$ & 0.2$\pm$0.0 $\quad$ & 0.2$\pm$0.0\\ \hline
Portugal / Data & 0.324$\pm$0.032 $\quad$& 0.331$\pm$0.006 $\quad$& 0.286$\pm$0.005\\
Portugal / Radiation & 4.639$\pm$1.881 $\quad$& 6.759$\pm$2.254 $\quad$& 198.3$\pm$186.4\\
Portugal / Gravity & 0.487$\pm$0.004 $\quad$& 0.527$\pm$0.005 $\quad$& 0.377$\pm$0.003\\
Portugal / Hierarchy & 0.258$\pm$0.000 $\quad$ & 0.258$\pm$0.000 $\quad$ & 0.258$\pm$0.000\\
Portugal /  Hierarchy-Admin  & 0.200$\pm$0.000 $\quad$ & 0.200$\pm$0.000   $\quad$ & 0.200$\pm$0.000 \\ \hline
France / Data & 0.196$\pm$0.007 $\quad$ & 0.290$\pm$0.081 $\quad$ & 0.154$\pm$0.004\\
France / Radiation & 13.60$\pm$5.42 $\quad$ & 656.8$\pm$507.2 $\quad$ & 25648$\pm$24798\\
France / Gravity & 0.287$\pm$0.003 $\quad$ & 0.263$\pm$0.002 $\quad$ & 0.166$\pm$0.002\\ 
France / Hierarchy & 0.158$\pm$0.000 $\quad$ & 0.158$\pm$0.000 $\quad$ & 0.158$\pm$0.000\\   
France /  Hierarchy-Admin  & 0.200$\pm$0.000 $\quad$ & 0.200$\pm$0.000   $\quad$ & 0.200$\pm$0.000  \\\hline
Country X / Data & 0.237$\pm$0.002 $\quad$& 0.168$\pm$0.002 $\quad$& 0.056$\pm$0.001\\
Country X / Radiation & 38.52$\pm$19.41 $\quad$& 9439$\pm$6096 $\quad$& 288.4$\pm$189.2\\
Country X / Gravity & 0.329$\pm$0.005 $\quad$& 0.286$\pm$0.009 $\quad$& 0.135$\pm$0.001\\
Country X / Hierarchy & 0.114$\pm$0.000 $\quad$ & 0.114$\pm$0.000 $\quad$ & 0.114$\pm$ 0.000\\ 
Country X /  Hierarchy-Admin &  0.200$\pm$0.000 $\quad$ & 0.200$\pm$0.000   $\quad$ & 0.200$\pm$0.000 \\ \hline
Ivory Coast / Data &  0.324$\pm$0.005  $\quad$ & 0.251$\pm$0.005  $\quad$ & 0.262$\pm$0.005\\
Ivory Coast / Radiation & 9.465$\pm$4.055 $\quad$ & 5.475$\pm$2.664 $\quad$ & 3.328$\pm$1.404\\
Ivory Coast / Gravity & 0.619$\pm$0.006 $\quad$ & 0.577$\pm$0.005 $\quad$ & 0.489$\pm$0.004\\
Ivory Coast / Hierarchy  & 0.255$\pm$0.000 $\quad$ & 0.255$\pm$0.000 $\quad$ & 0.255$\pm$0.000\\
Ivory Coast  /  Hierarchy-Admin &  0.200$\pm$0.000 $\quad$ & 0.200$\pm$0.000   $\quad$ & 0.200$\pm$0.000 \\
\end{tabular}
\end{center}
\end{table}

\begin{table}
\begin{center}
\caption{  {\bf (Table S3)} {\bf Benchmark measures quantifying the goodness of fit in Portugal, France, Country X and Ivory Coast.} \label{tab:benchmark}
The Dice (D), Sorensen (S), Cosine (C) and deviance (E) are four different measures of the distance between the actual and modeled networks. The correlation $corr$ measures a similarity between a model and the data. The parameters of the gravity and hierarchy models were chosen to minimize the value of E.}
\begin{tabular}{l | c | ccc | c | l}
    Country / Model & E$\times10^{-9}$ & D & S & C & corr & fitted parameters\\ 
    \hline
    Portugal / Radiation  & 314.1 &   0.781 & 0.739 & 0.476 &  0.525 &     \\     
    Portugal / Gravity & 79.80 & 0.865 & 0.419 & 0.844  & 0.145 &$\alpha=0.81$, $\beta =0.79$, $\gamma=-0.71$  \\ 
    Portugal / Hierarchy & 66.66 &0.346 & 0.404 & 0.308 & 0.683 & $q=0.258$ \\
    Portugal / Hierarchy-Admin & 74.20 &0.456 & 0.416 & 0.362 & 0.627 & $q=0.278$ \\
    \hline
    France / Radiation &  227.758 &0.618 & 0.647 & 0.270 & 0.730 &      \\     
    France / Gravity   & 90.905  & 0.267 & 0.524 & 0.185 & 0.815 &  $\alpha=0.69$, $\beta =0.69$, $\gamma=-1.44$ \\ 
    France / Hierarchy  & 73.524 &0.341 & 0.514 & 0.267 & 0.733 & $q=0.158$ \\
    France /  Hierarchy-Admin & 80.686 & 0.212 & 0.529 & 0.207 & 0.793 & $q=0.192$ \\
    \hline
    Country X / Radiation & 3.701 &  0.577 & 0.638 & 0.356  & 0.644 &    \\     
    Country X / Gravity   & 1.483 & 0.472 & 0.467 & 0.470 &   0.529  & $\alpha=0.81$, $\beta =0.78$, $\gamma=-1.06$ \\ 
    Country X / Hierarchy & 1.120 & 0.255 & 0.456 & 0.252 & 0.748 & $q=0.114$ \\
    Country X /  Hierarchy-Admin & 2.076 & 0.743 & 0.547 & 0.565 & 0.434 & $q=0.158$ \\
    \hline
    Ivory Coast / Radiation   &  268.18 & 0.701 & 0.703 & 0.358 & 0.645 &   \\     
    Ivory Coast / Gravity & 68.17 & 0.577 & 0.413 & 0.460 & 0.519 & $\alpha=0.94$, $\beta =0.94$, $\gamma =-0.51$  \\ 
    Ivory Coast / Hierarchy & 42.90  & 0.228 & 0.351 & 0.217 & 0.775 & $q=0.255$ \\
    Ivory Coast /  Hierarchy-Admin  & 65.98 &0.437 & 0.430 & 0.309 & 0.681 & $q=0.394$ \\
\end{tabular}
\end{center}
\end{table}

\begin{table*}
\caption{ {\bf (Table S4)} {\bf Over- / under-estimation measures of link at specific hierarchical distance in France, Portugal, Country X and Ivory Coast.} \label{tab:ratios}}
\begin{center}
\begin{tabular}{l | r r r r | }
    Country / Model  & $R_{h=1}$ & $R_{h=2}$& $R_{h=3}$ & $R_{h=4}$ \\ \hline
    UK / Gravity  &  0.54 & 0.73 & 1.15 &  1.33\\
    UK / Radiation  & 2.39 & 1.47 & 0.67 & 0.16  \\
    UK / Hierarchy & 1.10 & 0.73 & 0.90 & 1.18\\
    UK / Hierarchy-Admin & 0.25 &0.73 & 1.43 & 1.30\\\hline    
    Portugal /  Radiation    &  3.95 & 1.02 & 0.33 & 0.08 \\
    Portugal / Gravity   & 0.54 & 0.75 & 1.09 & 1.27 \\
    Portugal / Hierarchy   & 1.10 & 0.95 & 0.85 & 1.10\\   
    Portugal /  Hierarchy-Admin  & 0.70 & 1.14 & 1.17 & 0.93 \\  \hline     
    France / Radiation & 2.56 & 1.12 & 0.42 & 0.10  \\
    France / Gravity &  0.60 & 0.91 & 1.31  &  1.16\\
    France / Hierarchy  &  0.91 & 0.75 & 0.88  & 1.34 \\ 
    France / Hierarchy-Admin &  0.42 & 0.98 & 0.84 & 1.58 \\ \hline            
    Country X / Radiation   &  2.53 & 0.76 & 0.28 & 0.11 \\
    Country X / Gravity   & 0.81& 0.94 & 1.03& 0.52\\ 
    Country X / Hierarchy  & 1.25 & 0.59 & 0.58 & 2.09 \\ 
    Country X /  Radiation   &  3.95 & 1.02 & 0.33 & 0.08 \\
    Country X/  Hierarchy-Admin &0.12 & 0.43  & 1.59 & 2.60 \\  \hline        
    Ivory Coast / Radiation    & 4.51 & 1.79 & 0.63 & 0.12\\
    Ivory Coast / Gravity   & 0.29 & 0.47 & 1.02 & 1.33\\       
    Ivory Coast / Hierarchy   & 1.18 & 0.86 & 0.94 & 1.03 \\ 
    Ivory Coast / Hierarchy-Admin &  0.74 & 1.03  & 1.13  & 0.99  \\ 
\end{tabular}
\end{center}
\end{table*}

\clearpage

\section*{Supplementary Figures}

\begin{figure}[h!]
\begin{center}
\includegraphics{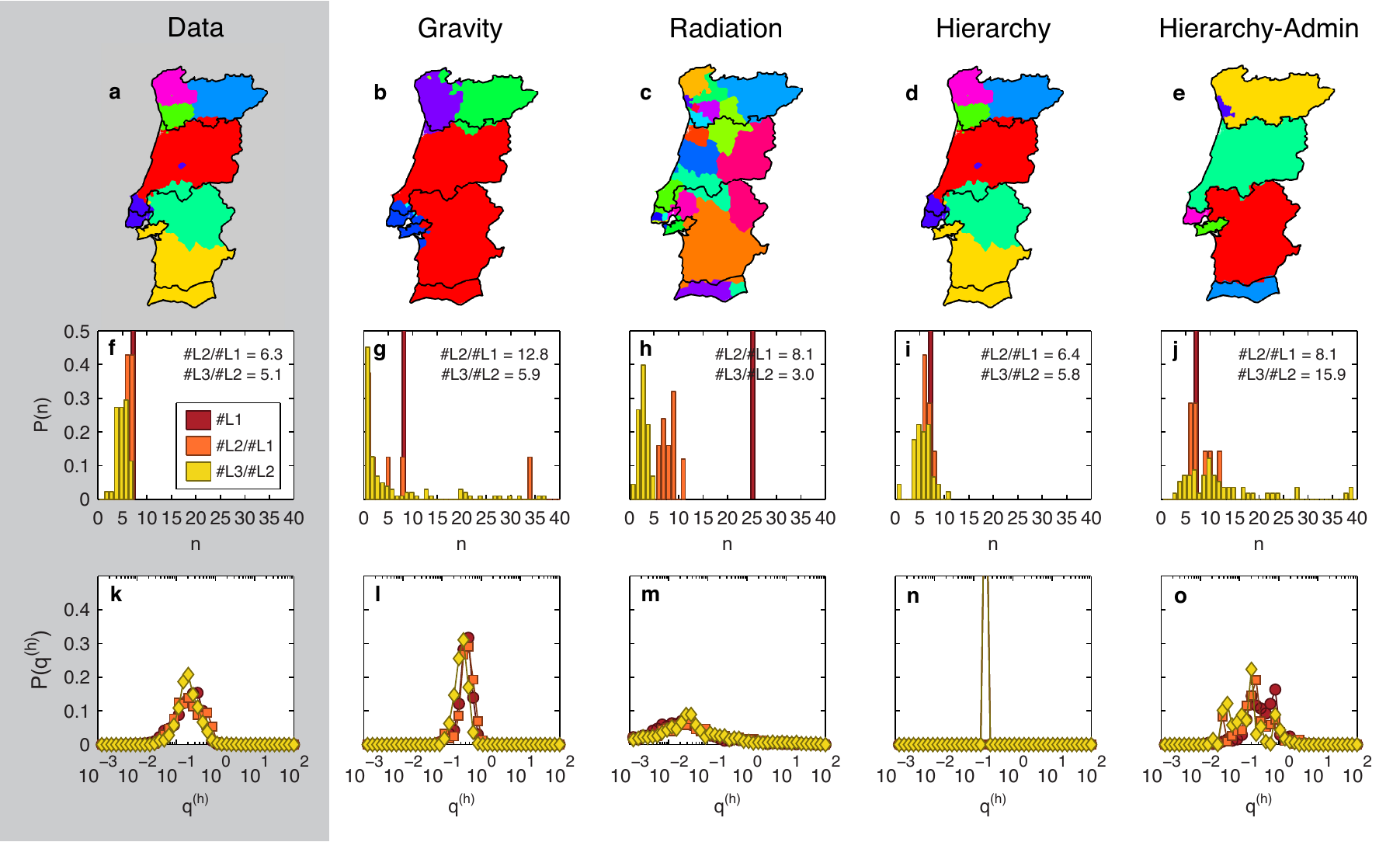}
\end{center} 
\caption{{\bf (Figure S1)} {\bf Hierarchical properties of spatial organization from human interactions in Portugal.} {\bf a--e}, Maps of $L_1$ communities in telephone call networks detected from data and from various interaction models. Black lines correspond to the administrative partitioning of the 5 NUTS1 regions of Portugal, colored areas to regions detected by a community detection algorithm applied to ({\bf a}) the data, and to the ({\bf b}) gravity, ({\bf c}) radiation, ({\bf d}) hierarchy, and ({\bf e}) administrative models. All detected regions are cohesive although some of the distinct colors used may appear similar.
{\bf f--j}, Probability distribution of number of subregions by region found in ({\bf f}) the actual network and ({\bf g--j}) in each model. The gravity model ({\bf g}) underestimates the number of $L_1$ communities but overestimates the numbers of subregions within regions. The radiation model ({\bf h}) strongly overestimates the number of $L_1$ communities. The hierarchy model ({\bf i}) correctly determines the distributions of sub-communities per community.
{\bf k--o}, Probability distributions of damping values $q^{(h)}$. The hierarchy model ({\bf n}) assumes a constant damping value for all levels.
\label{fig2PT}} 
\end{figure}

\begin{figure}
\begin{center}
\includegraphics{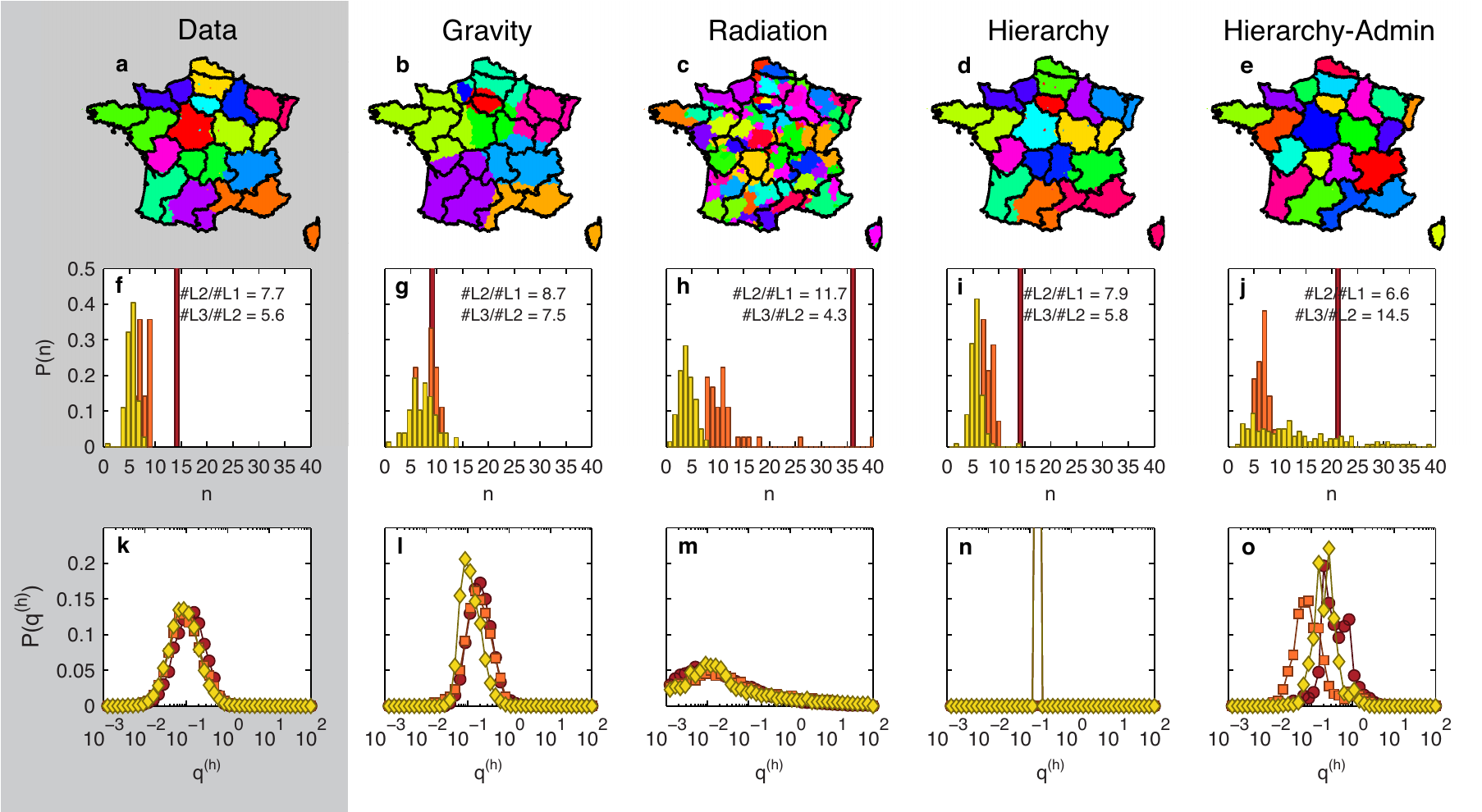}
\end{center} 
\caption{{\bf (Figure S2)} {\bf Hierarchical properties of spatial organization from human interactions in France.} {\bf a--e}, Maps of $L_1$ communities in telephone call networks detected from data and from various interaction models. Black lines correspond to the administrative partitioning of the 22 NUTS1 regions of France, colored areas to regions detected by a community detection algorithm applied to ({\bf a}) the data, and to the ({\bf b}) gravity, ({\bf c}) radiation, ({\bf d}) hierarchy, and ({\bf e}) administrative models. All detected regions are cohesive although some of the distinct colors used may appear similar.
{\bf f--j}, Probability distribution of number of subregions by region found in ({\bf f}) the actual network and ({\bf g--j}) in each model. The gravity model ({\bf g}) underestimates the number of $L_1$ communities but overestimates the numbers of subregions within regions. The radiation model ({\bf h}) strongly overestimates the number of $L_1$ communities. The hierarchy model ({\bf i}) correctly determines the distributions of sub-communities per community.
{\bf k--o}, Probability distributions of damping values $q^{(h)}$. The hierarchy model ({\bf n}) assumes a constant damping value for all levels.
\label{fig2FR}} 
\end{figure}

\begin{figure}
\begin{center}
\includegraphics{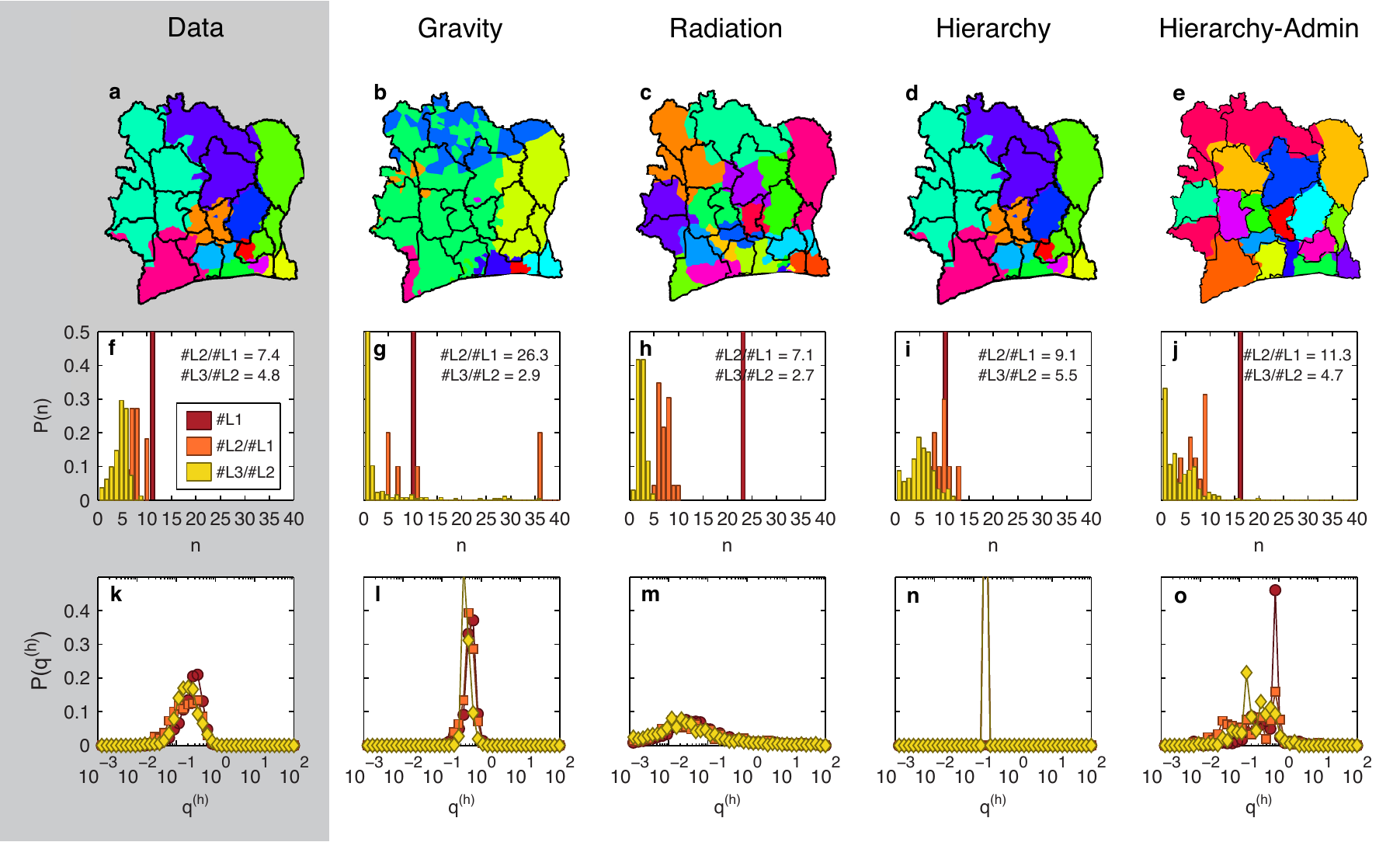}
\end{center} 
\caption{{\bf (Figure S3)} {\bf Hierarchical properties of spatial organization from human interactions in Ivory Coast.} {\bf a--e}, Maps of $L_1$ communities in telephone call networks detected from data and from various interaction models. Black lines correspond to the administrative partitioning of the 19 administrative regions of Ivory Coast, colored areas to regions detected by a community detection algorithm applied to ({\bf a}) the data, and to the ({\bf b}) gravity, ({\bf c}) radiation, ({\bf d}) hierarchy, and ({\bf e}) administrative models. All detected regions are cohesive although some of the distinct colors used may appear similar.
{\bf f--j}, Probability distribution of number of subregions by region found in ({\bf f}) the actual network and ({\bf g--j}) in each model. The gravity model ({\bf g}) underestimates the number of $L_1$ communities but overestimates the numbers of subregions within regions. The radiation model ({\bf h}) strongly overestimates the number of $L_1$ communities. The hierarchy model ({\bf i}) correctly determines the distributions of sub-communities per community.
{\bf k--o}, Probability distributions of damping values $q^{(h)}$. The hierarchy model ({\bf n}) assumes a constant damping value for all levels.
\label{fig2IC}} 
\end{figure}

\begin{figure}
\begin{center}
\includegraphics{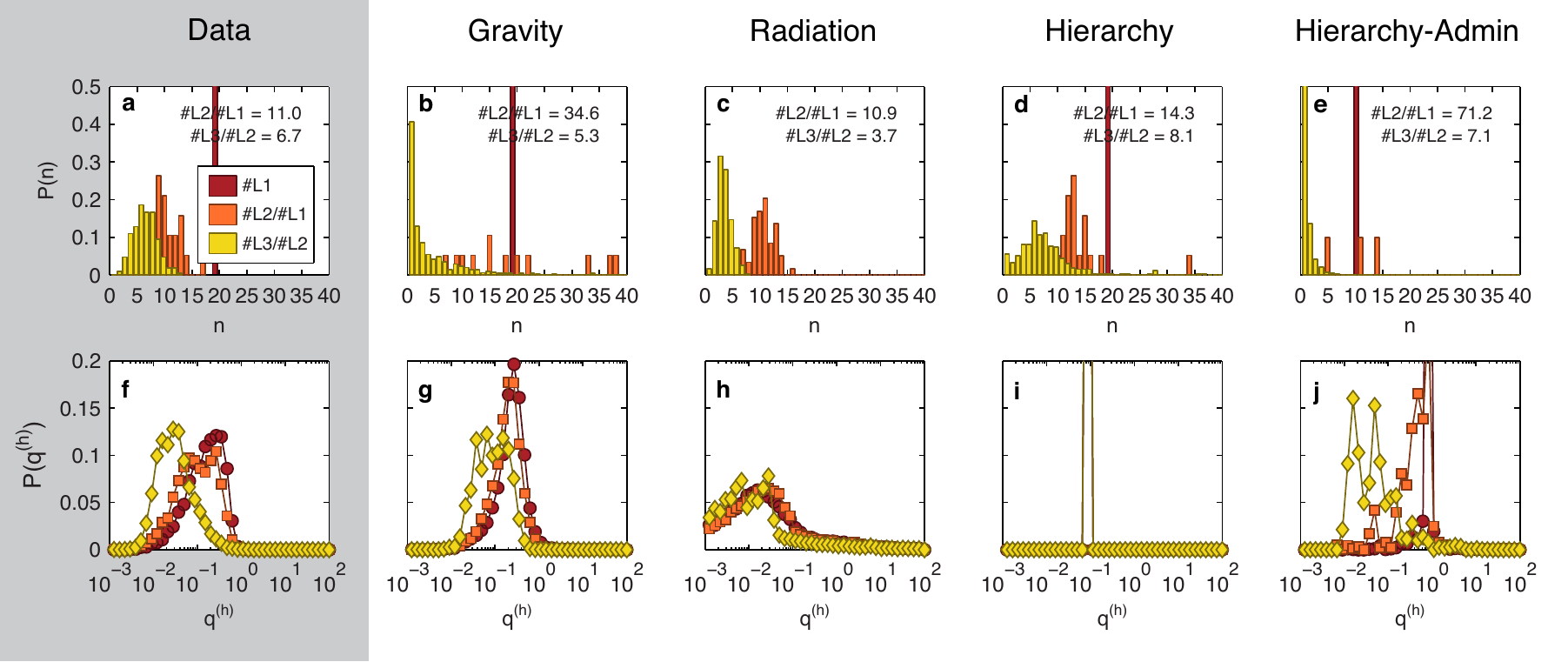}
\end{center} 
\caption{{\bf (Figure S4)} {\bf Hierarchical properties of spatial organization from human interactions in Country X.} {\bf a--e}, Probability distribution of number of subregions by region of Country X found in ({\bf a}) the actual network and ({\bf b--e}) in each model. The gravity model ({\bf b}) underestimates the number of $L_1$ communities but overestimates the numbers of subregions within regions. The radiation model ({\bf c}) strongly overestimates the number of $L_1$ communities. The hierarchy model ({\bf d}) correctly determines the distributions of sub-communities per community.
{\bf f--i}, Probability distributions of damping values $q^{(h)}$. The hierarchy model ({\bf h}) assumes a constant damping value for all levels. Maps of of $L_1$ communities are not shown as in other countries due to our non-disclosure agreement with the data providers from Country X. 
\label{fig2CX}} 
\end{figure}

\begin{figure}
\begin{center}
\includegraphics{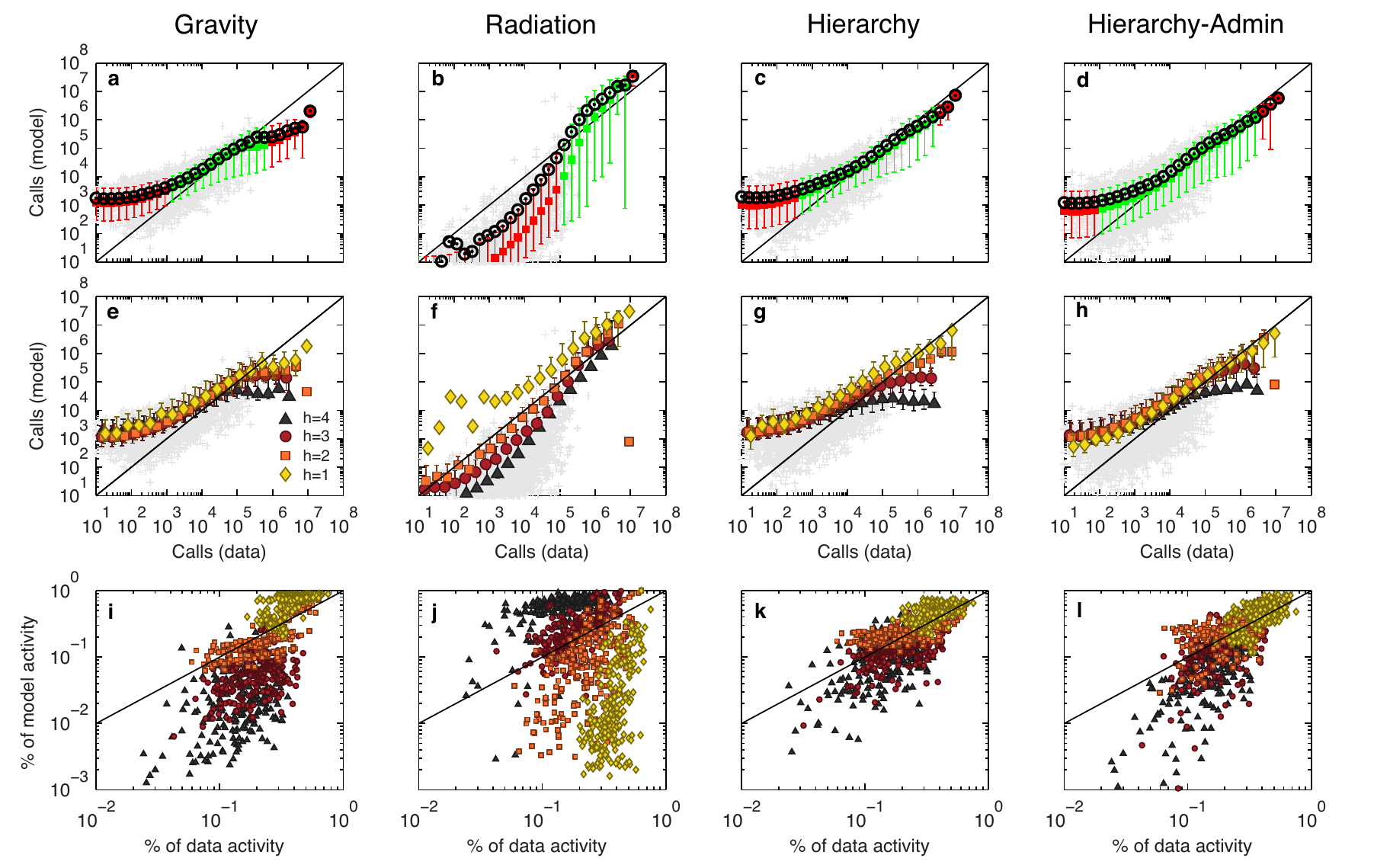}
\end{center} 
\caption{{\bf (Figure S5)} {\bf Comparison of model predictions in Portugal.} {\bf a--d}, Comparison of the actual total communication to the predicted communication for each pair of distinct locations, for the ({\bf a}) gravity, ({\bf b}) radiation, ({\bf c}) hierarchy, and ({\bf d}) administrative models. Gray markers are scatter plots for each pair of locations. A box is colored green if the equality line $y=x$ lies between the 9th and 91th percentiles in that bin and is red otherwise. Red boxes hence emphasize significant biases of the models. Black circles correspond to the average total communication of the pairs of locations in that bin.
{\bf e--h}, Goodness of prediction with respect to the hierarchical distance $h$, for the ({\bf e}) gravity, ({\bf f}) radiation, ({\bf g}) hierarchy, and ({\bf h}) administrative models. Gray markers are scatter plots for each pair of locations. Error bars show the corresponding 9th and 91th percentiles of total communication values. {\bf i--l}, For each L3 community, comparison of the fractions of activity of model versus data between that L3 community and L3 communities at different hierarchical distances, for the ({\bf i}) gravity, ({\bf j}) radiation, ({\bf k}) hierarchy and ({\bf l}) administrative models. 
\label{fig4PT}} 
\end{figure}

\begin{figure}
\begin{center}
\includegraphics[width=1.0\textwidth]{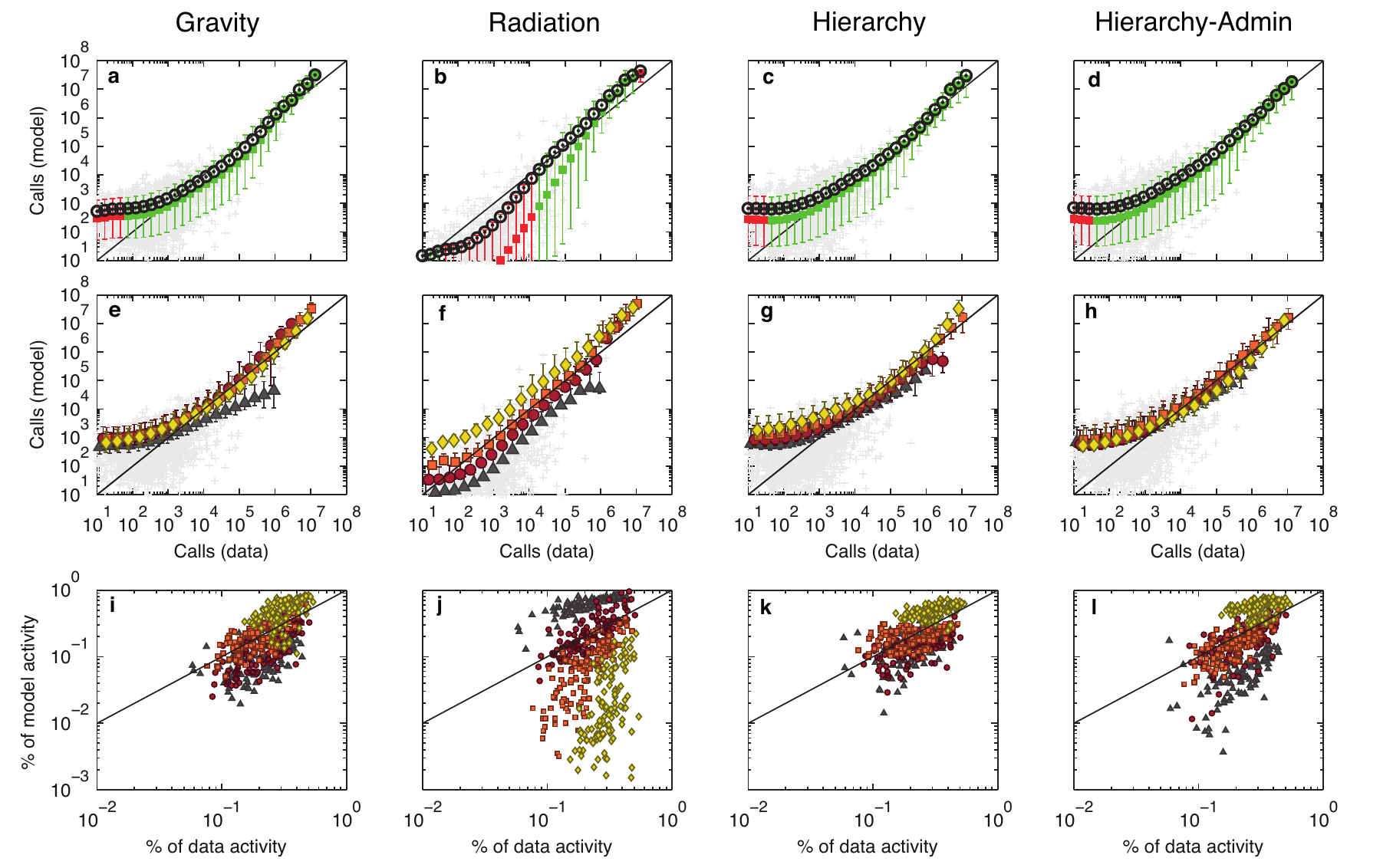}
\end{center} 
\caption{ {\bf (Figure S6)} {\bf Comparison of model predictions in France.} {\bf a--d}, Comparison of the actual total communication to the predicted communication for each pair of distinct locations, for the ({\bf a}) gravity, ({\bf b}) radiation, ({\bf c}) hierarchy, and ({\bf d}) administrative models. Gray markers are scatter plots for each pair of locations. A box is colored green if the equality line $y=x$ lies between the 9th and 91th percentiles in that bin and is red otherwise. Red boxes hence emphasize significant biases of the models. Black circles correspond to the average total communication of the pairs of locations in that bin.
{\bf e--h}, Goodness of prediction with respect to the hierarchical distance $h$, for the ({\bf e}) gravity, ({\bf f}) radiation, ({\bf g}) hierarchy, and ({\bf h}) administrative models. Gray markers are scatter plots for each pair of locations. Error bars show the corresponding 9th and 91th percentiles of total communication values. {\bf i--l}, For each L3 community, comparison of the fractions of activity of model versus data between that L3 community and L3 communities at different hierarchical distances, for the ({\bf i}) gravity, ({\bf j}) radiation, ({\bf k}) hierarchy and ({\bf l}) administrative models. 
\label{fig4FR}} 
\end{figure}

\begin{figure}
\begin{center}
\includegraphics{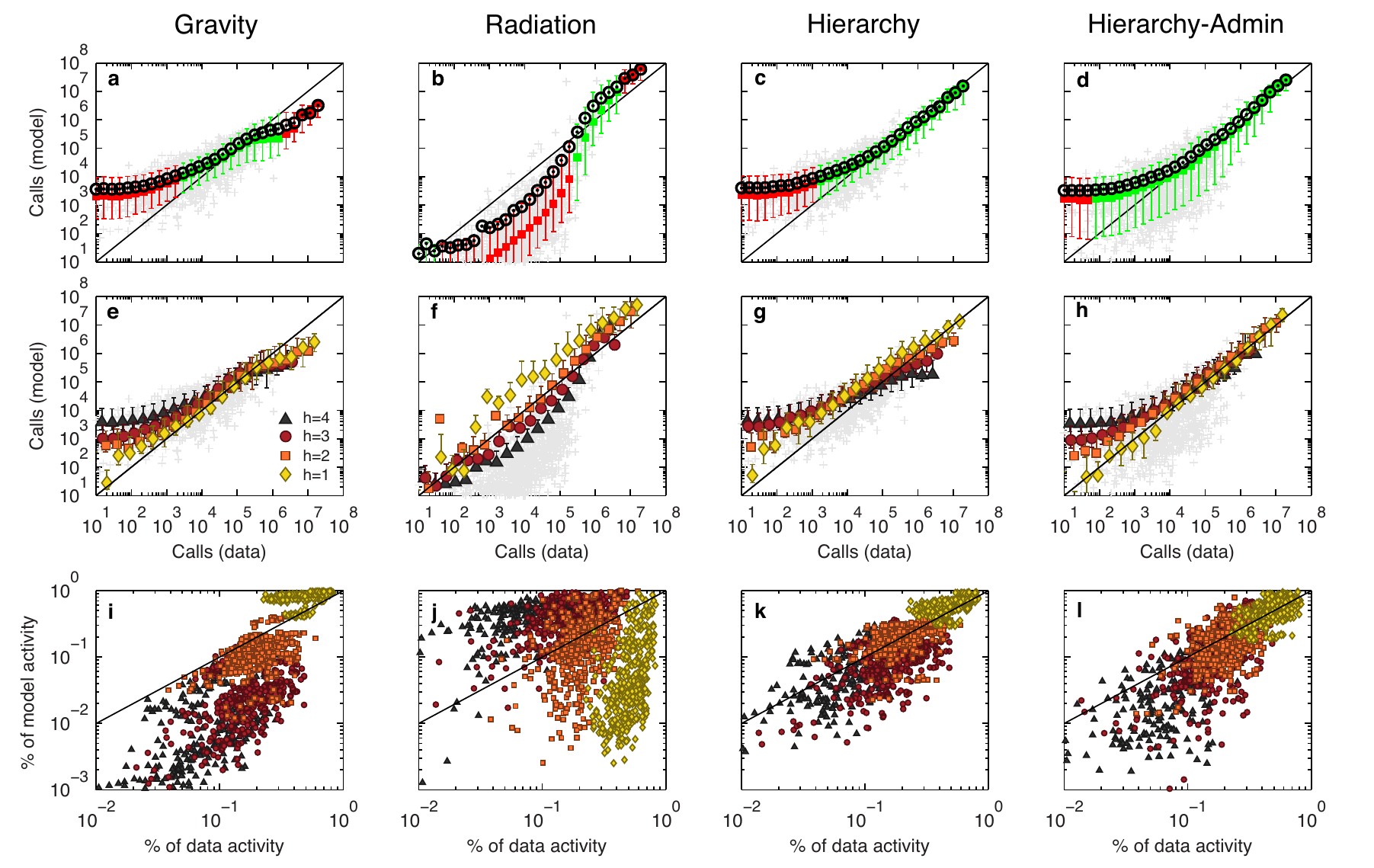}
\end{center} 
\caption{{\bf (Figure S7)} {\bf Comparison of model predictions in Ivory Coast.} {\bf a--d}, Comparison of the actual total communication to the predicted communication for each pair of distinct locations, for the ({\bf a}) gravity, ({\bf b}) radiation, ({\bf c}) hierarchy, and ({\bf d}) administrative models. Gray markers are scatter plots for each pair of locations. A box is colored green if the equality line $y=x$ lies between the 9th and 91th percentiles in that bin and is red otherwise. Red boxes hence emphasize significant biases of the models. Black circles correspond to the average total communication of the pairs of locations in that bin.
{\bf e--h}, Goodness of prediction with respect to the hierarchical distance $h$, for the ({\bf e}) gravity, ({\bf f}) radiation, ({\bf g}) hierarchy, and ({\bf h}) administrative models. Gray markers are scatter plots for each pair of locations. Error bars show the corresponding 9th and 91th percentiles of total communication values. {\bf i--l}, For each L3 community, comparison of the fractions of activity of model versus data between that L3 community and L3 communities at different hierarchical distances, for the ({\bf i}) gravity, ({\bf j}) radiation, ({\bf k}) hierarchy and ({\bf l}) administrative models. 
\label{fig4FR}} 
\end{figure}

\begin{figure}
\begin{center}
\includegraphics{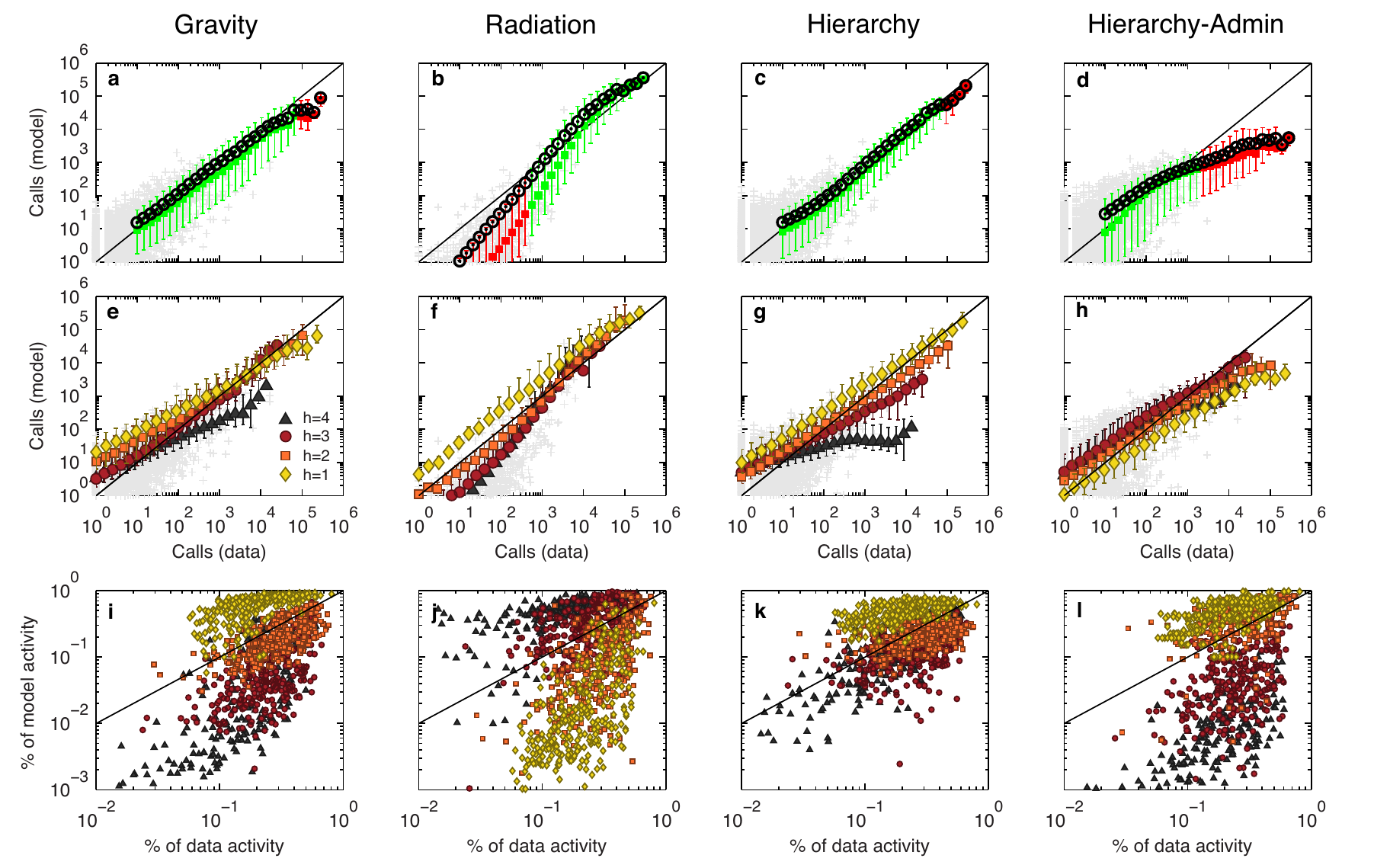}
\end{center} 
\caption{{\bf (Figure S8)} {\bf Comparison of model predictions in Country X.} {\bf a--d}, Comparison of the actual total communication to the predicted communication for each pair of distinct locations, for the ({\bf a}) gravity, ({\bf b}) radiation, ({\bf c}) hierarchy, and ({\bf d}) administrative models. Gray markers are scatter plots for each pair of locations. A box is colored green if the equality line $y=x$ lies between the 9th and 91th percentiles in that bin and is red otherwise. Red boxes hence emphasize significant biases of the models. Black circles correspond to the average total communication of the pairs of locations in that bin.
{\bf e--h}, Goodness of prediction with respect to the hierarchical distance $h$, for the ({\bf e}) gravity, ({\bf f}) radiation, ({\bf g}) hierarchy, and ({\bf h}) administrative models. Gray markers are scatter plots for each pair of locations. Error bars show the corresponding 9th and 91th percentiles of total communication values. {\bf i--l}, For each L3 community, comparison of the fractions of activity of model versus data between that L3 community and L3 communities at different hierarchical distances, for the ({\bf i}) gravity, ({\bf j}) radiation, ({\bf k}) hierarchy and ({\bf l}) administrative models. 
\label{fig4FR}} 
\end{figure}

\begin{figure*}[t!] 
\begin{center}
\includegraphics{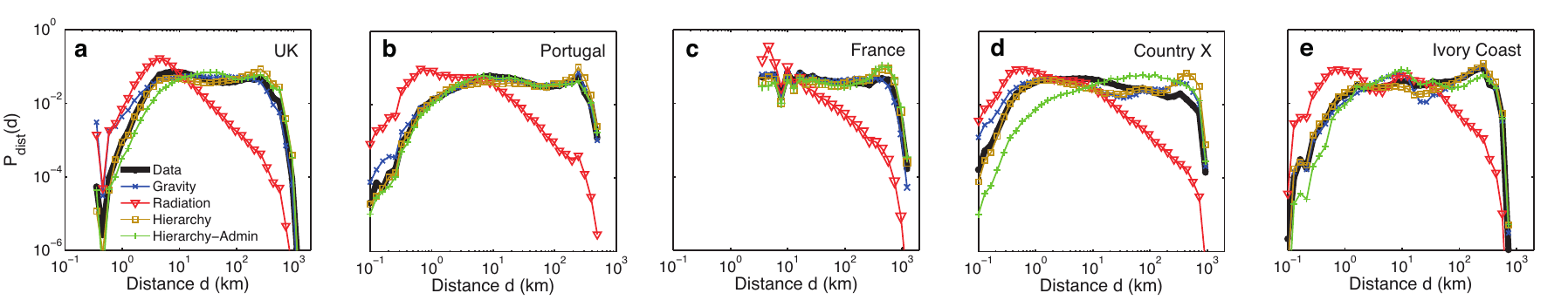}
\end{center} 
\caption{{\bf (Figure S9)} {\bf Comparing the model predictions.} The proportion  $P_{dist}(r)$ of communication occurring between two locations at a distance $r$ (in km) from each other, is measured in the data and in the models.
The radiation model is characterized by a lower than actual proportion of communication between distant (more than 100km) locations up to two orders of magnitude. It also presents a higher than actual proportion of communication between close (less than 10km with a peak between 0.5 and 5 km depending on the country) locations up to one order of magnitude. 
The gravity model presents in all countries a higher than actual proportion of communication between very close locations (100m-1km). In general, it also overestimates the low-range fluxes by 2 to 3 orders of magnitude and slightly underestimates the top-range fluxes.
The hierarchy model fits almost perfectly at low distances (less than 10km). Depending on the country, it only deviates slightly from the data at top-range fluxes or estimates them properly.
The fit of the hierarchy-admin model depends strongly on the country, but qualitatively comparable to the gravity and hierarchy models.
\label{comparing-dist}  } 
\end{figure*}

\begin{figure*}[t] 
\begin{center}
\includegraphics{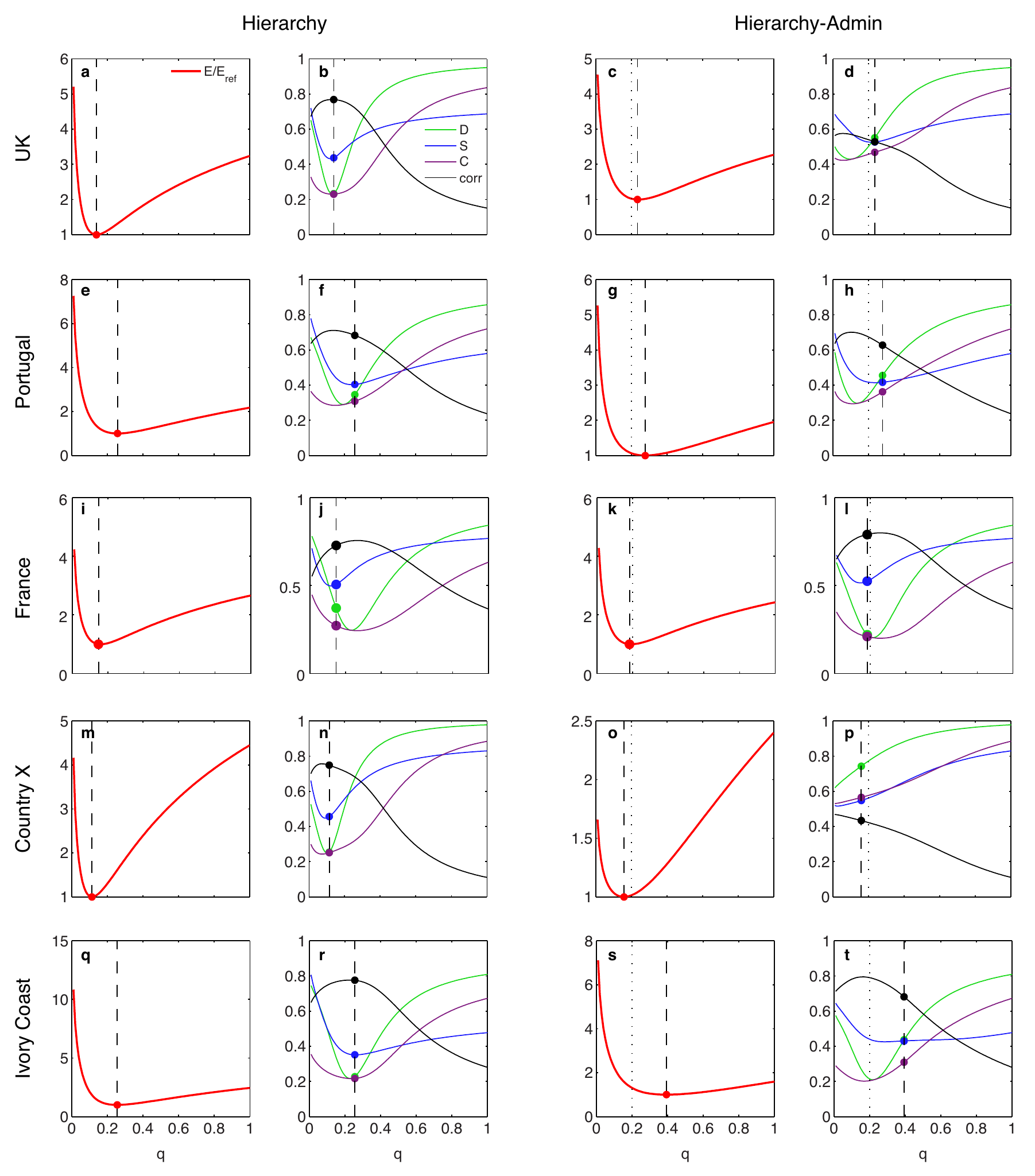}\\
\end{center} 
\caption{{\bf (Figure S10)} {\bf Stability analysis: benchmark measures of the Hierarchy and Hierarchy-Admin models with varying parameter $q$.} The dashed lines and circle markers show the `optimal' values of $q$ (reported in Supplementary Table 2) minimizing the deviance $E$. In all countries, either for the hierarchy or hierarchy-admin model, these optimal values of $q$ are also close to those minimizing the Dice ($D$), Sorensen ($S$) and Cosine ($C$) distances and maximizing the correlation ($corr$) between the data and the models. These optimal values are also stable, in the sense that close values of $q$ (roughly between $0.1$ and $0.3$) still provide benchmark measures close to their optimum. The dotted lines in the case of the Hierarchy-Admin model indicate the damping value $q=0.2$ matching robustly all countries. \label{qstability}} 
\end{figure*}

\clearpage

\section*{Supplementary Text}
\subsection*{Data partition versus administrative regions} 
\paragraph*{Partition overlap measures\\}
We use three classical measures of clustering similarity to quantify partition overlaps, i.e.~of how well two different partitions of the same set of locations match: Rand's criterion $\mathcal{R}$ \cite{rand1971}, Jaccard index $\mathcal{J}$ \cite{jain1988algorithms} and the Fowlkes and Mallows index $\mathcal{F}$ \cite{Fowlkes:1983vq}. 
Consider two partitions $\mathcal{C}$ and $\mathcal{C}'$ of a set of $n$ nodes and note
\begin{itemize} 
\item $n_{11}$ the number of pairs of nodes in the same community both in $\mathcal{C}$ and $\mathcal{C}'$; 
\item $n_{01}$ the number of pairs of nodes not in the same community in $\mathcal{C}$ but in the same community in $\mathcal{C}'$; 
\item $n_{10}$ the number of pairs of nodes in the same community in $\mathcal{C}$ but not in the same community in $\mathcal{C}'$; 
\item $n_{00}$ the number of pairs of nodes in different communities both in $\mathcal{C}$ and $\mathcal{C}'$.
\end{itemize} 
All types of pairs being taken into account, we have $n_{11}+n_{01}+n_{10}+n_{00}=n(n-1)/2$. The  $\mathcal{R}$,  $\mathcal{J}$ and $\mathcal{F}$ indices are then defined by: 
\begin{equation}
\mathcal{R} =\frac{2(n_{11}+n_{00})}{n(n-1)},  \quad \quad  
\mathcal{J} =\frac{n_{11}}{n_{11}+n_{10}+n_{01}},  \quad \quad  
\mathcal{F} =\frac{n_{11}}{\sqrt{(n_{11}+n_{10})(n_{11}+n_{01})}} 
\end{equation}
which are different ways of quantifying how well the partitions match pairs of nodes. The values of $\mathcal{R},\mathcal{J},\mathcal{F}$ lie between 0 and 1, and values close to 1 indicate a perfect match between the two compared partitions. However, even for the case of two completely unrelated clusterings, all indices are in general strictly larger than zero, more so for $\mathcal{R}$ \cite{Fowlkes:1983vq}. Therefore, to have a baseline, we calculated the average indices over 10000 random shufflings of the nodes's clusters, denoted by $\bar{\mathcal{R}_\mathrm{r}}$, $\bar{\mathcal{J}_\mathrm{r}}$ and $\bar{\mathcal{F}_\mathrm{r}}$ (see Table \ref{tab:regions}). 

\vspace{3mm}
\paragraph*{Results\\}
The partitions obtain for levels $L_1$, $L_2$ and $L_3$ correspond in general to geographically cohesive regions, and are rather similar to administrative regions in number and size. As previously noted in \cite{ratti2010rmg,sobolevsky2013delineating,sobolevsky2013general} (from which the partitions resulting from the Combo algorithm for UK and Portugal are reproduced), these results may come as a surprise regarding that the modularity approach of the Combo algorithm has no spatial constraint nor does it impose any restriction on the number of communities.

The partition overlap measures given in Table \ref{tab:regions}, together with p-values assessing the statistical significance of the partition with respect to a null model in which communities of cell towers were randomly reshuffled, quantitatively confirm the similarity between the administrative partitions (we choose to refer to the european {\em Nomenclature of Territorial Units for Statistics} - or NUTS - standard for the european countries) and the Combo partition. For example, the $L_1$ UK partitioning shows values of $\mathcal{R}=0.954$ with a baseline $\bar{\mathcal{R}_\mathrm{r}} = 0.825$, $\mathcal{J}=0.618$ with a baseline $\bar{\mathcal{J}_\mathrm{r}} = 0.049$, and $\mathcal{F}=0.769$ with a baseline $\bar{\mathcal{F}_\mathrm{r}} = 0.096$ - with all significance measures ($p<10^{-4}$) indicating a good match between the administrative and the Combo partitions. Going a step further and comparing the $L_2$ and $L_3$ communities with NUTS regions of corresponding levels, the match between the Combo and administrative partitions is still good, as indicated by significantly higher than average overlap measures. The $L_2$ UK partitioning hence shows values of $\mathcal{R}=0.972$ with a baseline $\bar{\mathcal{R}_\mathrm{r}} = 0.957$, $\mathcal{J}=0.222$ with a baseline $\bar{\mathcal{J}_\mathrm{r}} = 0.007$, and $\mathcal{F}=0.439$ with a baseline $\bar{\mathcal{F}_\mathrm{r}} = 0.018$ and the $L_3$ UK partitioning shows values of $\mathcal{R}=0.988$ with a baseline $\bar{\mathcal{R}_\mathrm{r}} = 0.986$, $\mathcal{J}=0.099$ with a baseline $\bar{\mathcal{J}_\mathrm{r}} = 0.001$, and $\mathcal{F}=0.292$ with a baseline $\bar{\mathcal{F}_\mathrm{r}} = 0.004$. Note that while all these values stay significant, the differences between the Combo / administrative overlap values and the random / administrative overlap values decrease when we look at more fine-grained partitions. This has to be expected since the deviations between the Combo and administrative partition can only increase with the level of partition as the deviations at a given level automatically impact the successive levels. Similar results can be drawn from the other countries, see Table \ref{tab:regions}. 

Table \ref{tab:regions} also display the modularity scores $Q^*_{Combo}$ and $Q^*_{off}$ of the Combo and administrative partitions at the different levels. To be consistent with the procedure of the Combo algorithm, which builds the level $n+1$ subpartition by decomposing each community of the level $n$ partition, these modularity scores indicate in case of $L_2$  (resp. $L_3$) partitions the average modularity score of each subpartition defined with respect to a subnetwork inside each corresponding $L_1$ (resp. $L_2$) community. Our measures show that the Combo partition always has a better modularity score than the administrative partition. The modularity scores also decrease when we look at higher level partitions, indicating that the $L_1$ communities are the most relevant.

\vspace{3mm}
\paragraph*{Interpretation\\}
The similarity between the regions emerging from the communication network through the Combo procedure and the administrative boundaries can be interpreted as a natural evidence towards the latter's validity \cite{ratti2010rmg,sobolevsky2013delineating}. In view of the deviation between the two partitions, Combo partitioning appears to be more aligned with human interactions, as measured by the modularity score, suggesting that the border of administrative regions sometimes deviates from the underlying reality of interactions. 
Most interestingly, the partitioning created by our approach provides a unified hierarchical framework to compare the geographical structure of human interactions in different countries, which present an important alternative to the administrative boundaries whose shape and number depend substantially on the historical and political context of each country as well as particular local regional delineation policy.

\begin{table}[h!]
\begin{center} 
\caption{ {\bf Overlap between the administrative regions and the community found by the Combo algorithm.} \label{tab:regions}}
\begin{tabular}{c c | r r |  r r |  r r r  } 
	Country & Network & $N_{off}$ & $N_{combo}$ & $Q^*_{off}$ & $Q^*_{Combo}$ & $ \mathcal{R}$($\bar{\mathcal{R}_\mathrm{r}}$) & $\mathcal{J}$($\bar{\mathcal{J}_\mathrm{r}}$)  &$\mathcal{F}$($\bar{\mathcal{F}_\mathrm{r}}$)\\ 
	\hline	
	UK & NUTS1/$L_1$ & 11 & 16 & 0.642 & 0.657 & 0.954(0.825) & 0.618(0.049) & 0.769(0.096)\\ 
	UK & NUTS2/$L_2$ & 36 & 150 & 0.490 & 0.631 & 0.972(0.957) & 0.222(0.007) & 0.439(0.018)\\ 
	UK & NUTS3/$L_3$ & 133 & 917 & 0.415 & 0.472 & 0.988(0.986) & 0.099(0.001) & 0.292(0.004)\\ 
	\hline
	Portugal & NUTS2/$L_1$ & 5  &  7  & 0.445 & 0.491 & 0.859(0.669) & 0.496(0.113) & 0.671(0.206)\\ 
	Portugal & NUTS3/$L_2$ & 28 & 44 & 0.358 & 0.478 & 0.935(0.881) & 0.314(0.027) & 0.523(0.057) \\
	Portugal & NUTS4/$L_3$ & 275 & 226 & 0.359 & 0.361 & 0.984(0.975) & 0.230(0.005) & 0.411(0.011) \\	
	\hline	
	France & NUTS1/$L_1$ & 22 & 14 & 0.638 & 0.645 & 0.964(0.874) & 0.579(0.033) & 0.748(0.066)  \\ 
	France & NUTS2/$L_2$ & 96 & 108 & 0.486 & 0.592 &  0.994(0.978) & 0.609(0.006) & 0.758(0.011) \\ 
	France & NUTS3/$L_3$ & 335 & 609 & 0.347 & 0.405 &0.997(0.994) & 0.336(0.001) & 0.522(0.003) \\
	\hline 
	 Country X & NUTS1/$L_1$ & - & 11 & 0.420 & 0.453 & 0.862(0.756) & 0.342(0.073) & 0.518(0.139) \\
	 Country X & NUTS2/$L_2$ & - & 81 & 0.312 & 0.480 & 0.874(0.843) & 0.127(0.018) & 0.338(0.054)\\
	 Country X & NUTS3/$L_3$ & - & 388 & 0.315 & 0.341 &0.907(0.901) & 0.031(0.003) & 0.160(0.018)\\ 
	\hline
	 Ivory Coast & regions/$L_1$ & 19 & 19 & 0.666 & 0.807 & 0.845(0.761) & 0.264(0.055) & 0.474(0.118) \\
	 Ivory Coast & departments/$L_2$ & 50 & 209 & 0.306 & 0.589 & 0.900(0.881) & 0.097(0.010) & 0.308(0.035) \\
	 Ivory Coast & prefectures/$L_3$ & 255 & 1401 & 0.277 & 0.452 & 0.966(0.962) & 0.057(0.002) & 0.236(0.009)\\ 	 
\end{tabular}
\end{center} 
\begin{flushleft}
{\footnotesize
The ``Network'' column indicates the levels of the administrative partition and of the community partition that are compared \cite{sobolevsky2013delineating}. Columns $N_{off}$ and $N_{combo}$ respectively refer to the number of NUTS regions and the number of communities found by the community detection algorithm at the considered level. Columns $Q^*_{off}$ and $Q^*_{Combo}$ indicate the average modularity score of all the administrative or Combo sub-partitions at the considered level.  
$\bar{\mathcal{R}_\mathrm{r}}$, $\bar{\mathcal{J}_\mathrm{r}}$ and $\bar{\mathcal{F}_\mathrm{r}}$ give the baselines for RandÕs criterion $\mathcal{R}$, the Jaccard index $\mathcal{J}$ and the Fowlkes-Mallows index $\mathcal{F}$. The closer $\mathcal{R}$, $\mathcal{J}$ or $\mathcal{F}$ to 1, the better the overlap of the detected communities with the administrative regions. The baseline values of the similarity indices are computed on 10000 random shuffling of the nodes's clusters. For the 3 used measures, none of these random shufflings is more similar to the administrative partition than the partition found by the community detection algorithm: the statistical significances of all the similarities have a p-value $< 10^{-4}$. }
\end{flushleft}
\end{table}

\subsection*{Central Place Theory}

Classical theories of urban planning have traditionally suggested economical and geographic laws to systematically determine the arrangement of towns and cities. In particular, the central place theory (CPT) developed by Christaller \cite{christaller1933zentralen,christaller1966central} seeks to explain the number, size and location of human settlements in an urban system. The basic assumption of the theory is that settlements function as `central places' providing services to surrounding areas. The hierarchy of the cities is based on the range of goods and services they provide. Low order goods and services (groceries, bakeries, post offices) are present in all places, including small and large centers. Higher order goods and services (jewellery, large malls, universities) are only present in large centers, which are less numerous. These centers are supported by a large population, including its own and those of the surrounding smaller centers. The lowest settlements should form an hexagonal lattice, this being the most efficient regular pattern to serve areas without any overlap (in terms of radial distance and perimeter for a fixed area). Settlements of higher order (villages, cities) should then be regularly spaced on an hexagonal pattern of higher radius, with their centers placed on centers of hexagons of the lowest order. 

Christaller defined $K$-hexagonal landscapes as arrangements where each higher order settlement is supported by $K-1$ lower order settlements and itself. Christaller and later L\"osch \cite{losch1944raumliche} both developed arguments over which value of $K$ is adapted to describe different situations. For example, $K=3$ is suited for sharing local goods (marketing principle), $K = 4$ is suited for reducing cost of transport (traffic principle) while $K = 7$ - a case where each satellite depends only on one center - is suited for political stability (administrative principle). 
Christaller conceived these models as hierarchical, with all higher order places in the hexagon surrounded by lower order places to explain not only local but regional economics and spatialization of urban centers, the value of $K$ possibly changing from level to level. 

Filling the gap between CPT and reality, several distortions to the original model have been introduced over time to account for inhomogeneous population densities, resource locations, or specialization of cities. However, nowadays CPT is not a part of modern regional science and it has been criticized for being a static theory, not explaining how central places emerge and develop \cite{veneris1984inf}. It has also been shown that while CPT is ideally suited to describe agricultural areas, it is less relevant to describe industrial areas which are highly diversified in nature.  
But despite its imperfections, CPT still remains one of the strongest economic theories for understanding the spatial organization of the society as the hierarchy of urban centers. It has been applied in regional and urban economies, describing the location of trade and service activity, and for describing consumer market-oriented manufacturing.

\subsection*{Models' variations} 

In parallel with the models presented in the main text, we also tested different variations of the gravity, radiation and hierarchy models to predict human interactivity. In the following section, we present the definition of these variations and standard benchmark measures quantifying how well these models fit the data.
As for the models presented in the main text, we always use the total amount of communication $w_i$ originating from a location $i$ as a proxy for its population.

\subsubsection*{Definitions}
 
\paragraph*{Radiation model versions} 
The radiation model is a parameter-free model recently introduced in the context of migration patterns \cite{simini2012umm}. Given the distance $d_{ij}$ between two locations $i$ and $j$, the model predicts that the flux of individual moves $T_{ij}$ between those two locations will depend on the population at the origin, the population at the destination and on the population $s_{ij}$ within the circle of radius $d_{ij}$ centered on the origin location $i$. Applied to our case, the radiation model is written as 
\begin{equation} 
T_{ij}^{Rad}  = C_i \frac{w_i w_j}{(w_i + s_{ij})(w_i + w_j+ s_{ij})} =  C_i w_i \left(  \frac{1}{w_i + s_{ij}} - \frac{1}{w_i + w_j+  s_{ij}} \right),
\label{EQ:RM} 
\end{equation} 
where $s_{ij}=\sum_{k,\, 0<d_{ik}<d_{ij}}w_k$ is the total amount of communication originating from locations at a distance shorter than $d_{ij}$ from location $i$ and $C_i$ is a normalization factor ensuring that the predicted total activity of each node is the same as the actual one, i.e. $\sum_{j\neq i} T_{ij}^{Rad}= \sum_{j\neq i} T_{ij}$. The model is otherwise parameter-free.

We also applied the generalized version of the radiation model proposed in \cite{simini2013human}, introducing a parameter $\lambda$ which can be interpreted in our case as a fraction of individuals people will not consider as potential contacts because of lack of information about them of other reasons. This version of the radiation model can be written as     
\begin{equation}
T_{ij}^{Gen Rad}  = C_i\, w_i \, \frac{1}{1-\lambda}  \left(  \frac{1-\lambda^{w_i+s_{ij}}}{w_i + s_{ij}} -    \frac{1-\lambda^{w_i+w_j+s_{ij}}}{w_i + w_j+  s_{ij}} \right).
\label{EQ:NRM} 
\end{equation}
where $C_i$ is again a normalization factor ensuring that $\sum_{j\neq i} T_{ij}^{Gen Rad}= \sum_{j\neq i} T_{ij}$. Notice that the case $\lambda = 0$ corresponds to the original radiation model.

\vspace{1cm}
\paragraph*{Gravity models} 
The gravity model is one of the oldest models describing human mobility and interaction, formulated in analogy to Newton's law of gravity. Here interaction strength or mobility fluxes between a source $i$ and a destination $j$ are originally proposed to be related to a function of the distance $d_{ij}$ between the two locations and the product of the (powers of) population at the source an at the destination, $T_{ij}^{Grav} \sim w_i^\alpha w_j^\beta f(d_{ij})$. 
We tested the two classical forms of the gravity model using a power or an exponential law as deterrence function: 
\begin{eqnarray}
T_{ij}^{Grav PL}  &=&  C w_i^{\alpha} w_j^{\beta} d_{ij}^{\gamma}, 	\label{EQ:GplM} \\
T_{ij}^{Grav EXP} &=& C w_i^{\alpha}w_j^{\beta} \exp( -d_{ij}/d_0)	\label{EQ:GexpM} 
\end{eqnarray} 
where in both cases $C$ is a global normalization constant ensuring that $\sum_{i,\, j\neq i} T_{ij}^{Grav}= \sum_{i,\,j\neq i} T_{ij}$ and $\alpha$, $\beta$, $\gamma$, $d_0$ are parameters to fit. For example, taking the logarithm on both sides of Eq. (\ref{EQ:GplM}), we obtain $\log{T^{Grav PL}_{ij}} = \log{C} + \alpha\log{w_i} + \beta \log{w_{j}}  + \gamma\log{d_{ij}}$. We then use, as is customary, $\log{T^{Grav PL}_{ij}}$ to estimate the different parameters through a regression analysis \cite{flowerdew1982method}. We selected the set of parameters that minimized the deviance $E$ (see definition in main text, Methods section).
We also computed versions of these models where the population exponents are fixed, i.e. $\alpha = \beta = 1$.

At this point, we observe an often overseen difference between the gravity model relying on a single global normalization factor $C$ and the radiation model relying on local normalization factors $\{C_i\}$ - one for each location. One could argue that this difference gives an advantage to the radiation model. For the sake of comparison, we then also tested {\em locally constrained } gravity models that can be written as:  
\begin{eqnarray}
T_{ij}^{Grav PL\,loc} &=& C_i w_i w_j d_{ij}^{\gamma} \label{EQ:GplMloc}  \\ 
T_{ij}^{Grav EXP\,loc} &=& C_i w_i w_j \exp(-d_{ij}/d_0) \label{EQ:GexpMloc} 
\end{eqnarray} 
where in each case $C_i$ is a normalization factor ensuring that the predicted total activity of each node is the same as the actual one. This type of constrained model was already presented in \cite{wilson1967statistical} and more recently in \cite{sagarra2013statistical}.

%
\vspace{1cm}
\paragraph*{Hierarchical models} 
The general idea behind the hierarchical model is to simply replace the actual distance used in the gravity models by a hierarchical distance. In its most generic definition, the hierarchical model predicts an interaction strength between location $i$ and $j$ to be written as $T_{ij}^{Hier} = C_i w_i^\alpha w_j^\beta f(h_{ij})$, where $h_{ij}$ is the hierarchical distance between these two locations based on the Combo partition or any other partition and $f$ is a deterrence function. As it is done for the gravity model, one could a priori choose any deterrence function.  
We tested different simple forms of hierarchy models, using the hierarchical distances $\{h_{ij}\}$ provided either by the Combo partition or the administrative partition. Hierarchy models using a global normalization framework - as the usual gravity models do - read:
\begin{eqnarray}
T_{ij}^{Hier EXP} &=& C w_i^{\alpha} w_j^{\beta} q^{h_{ij}} \label{EQ:HexpM}\\
T_{ij}^{Hier PL}  &=&  C w_i^{\alpha} w_j^{\beta} h_{ij}^{\gamma}, \label{EQ:HplM}
\end{eqnarray} 
where $C$ is a global normalization constant, and $\alpha$, $\beta$, $\gamma$ and $q$ are parameters which we fit by minimizing the deviance $E$, see below. 
We also computed versions of these models where the population exponents are fixed, i.e. $\alpha = \beta = 1$.

The locally constrained versions of models given in Eqs. (\ref{EQ:HexpM}) and (\ref{EQ:HplM}) with fixed population exponent read:
\begin{eqnarray}
T_{ij}^{Hier EXP\,loc} &=& C_i w_i w_j q^{h_{ij}} \label{EQ:HexpMloc}\\
T_{ij}^{Hier PL\,loc}  &=&  C_i w_i w_j h_{ij}^{\gamma}, \label{EQ:HplMloc}
\end{eqnarray} 

It turns out that the model given by Eq (\ref{EQ:HexpMloc}) is naturally related to the notion of damping parameter. Indeed, assuming $\beta=1$ in the generic functional form $T_{ij}^{Hier} = C_i w_i^\alpha w_j^\beta f(h_{ij})$ and taking into account that the normalization factor $C_i$ ensures $w_i^{Hier} = \sum_j T_{ij}^{Hier (h)}=w_i$ implies
\begin{eqnarray*}
T_{i}^{Hier (h)} &=& \sum_{j,\,h_{ij}=h}T_{ij}^{Hier}\\ 
&=&  C_i w_i^\alpha f(h) \sum_{j,\,h_{ij}=h} w_j \\ 
&=&  C_i w_i^\alpha f(h) \sum_{j,\,h_{ij}=h} w_j^{Hier (h)}  \\
&=& C_i w_i^\alpha f(h) W_{i}^{Hier (h)} 
\end{eqnarray*} 
and thus
\begin{eqnarray}
q_{i}^{Hier (h)} =   \frac{T^{Hier (h+1)}_{i}}{T^{Hier (h)}_{i}} \frac{W^{Hier (h)}_i}{W^{Hier (h+1)}_i} = \frac{f(h+1)}{f(h)}, 
\end{eqnarray} 
an equation which immediately implies that choosing an exponential deterrence function $f(h)=q^h$ will ensure a constant damping parameter with respect to the locations and hierarchy levels $q_{i}^{Hier (h)} = q$ for all $i,h$.

%
\vspace{1cm}
\paragraph*{Hierarchical-Admin models} 
The hierarchical models rely on the notion of hierarchical distances between locations, which depend on a given partition. For cases when the communication network or it's partitioning are not known, we can as well defined a model based on Administrative partitions of the countries.
In the following, we refer to hierarchy models based on Administrative partition as `hierarchy-admin' models.


\subsubsection*{Analysis} 
Benchmark measures (as defined in main text) of the different models along with their fitted parameters are reported in Tables \ref{table:benchmark} and \ref{table:benchmark2}. We make a number of remarkable observations:

\begin{itemize}
\item In every country and according to all benchmark measures, the generalized radiation model is significantly more appropriate than the original one to describe our communication networks (e.g. in UK, the cosine distance to the data goes from $0.344$ in the original radiation model to $0.195$ in the generalized radiation model, similarly the correlation to the data goes from $0.656$ to $0.805$).
\item Locally constrained models always perform better compared to their `global normalization' counterpart. 
\item When using an exponential deterrence function, the hierarchy models based on our Combo partition are always significantly better than the corresponding gravity models, with respect to the deviance or any other benchmark measure. E.g. in UK, the cosine distance goes from $0.595$ in the `gravity EXP' model to $0.278$ in the `Hierarchy EXP' model, similarly the correlation to the data goes from $0.402$ to $0.72$. The Hierarchy models using the administrative partition are also slightly better than the corresponding gravity models, except in Country X where we already observed some bias due to the administrative borders (see section II-D above). 
\item When using a power law deterrence function, the best model can vary with respect to the benchmark measure and the studied country. The `Gravity PL loc' model is in general the best one.   
\item As one could expect, the models where the population exponent $\alpha$ and $\beta$ are not constrained always have a lower deviance $E$ than the corresponding models with $\alpha=\beta=1$ (but be aware that a correct comparison of the performance of the models based on the deviance should also take the number of parameter into account). According to the other benchmark measures (not used for the fitting), the unconstrained models are also most often better fit than the constrained ones (e.g. their correlation to data is higher $12$ times out of $18$) and when it's not the case the difference between the fit measures is small.
\item In general, the gravity PL model is better than the gravity EXP model. It is the best one regarding all benchmark measures in UK and Country X, and two out of five measures in Portugal.
\item The hierarchy EXP models always fit better than the hierarchy PL models, all measures and countries considered.
\end{itemize}

The 20 tested models can be ranked according to the average of the benchmark measures taken over all countries, see Table \ref{tab:rankModel}. 
While on average the best model is always the constrained `Gravity PL loc' model, the `hierarchy EXP loc' model presented in the main text is a close second (and first in some countries). 7 of the top 10 models are versions of the hierarchy model. In particular, the `hierarchy-admin EXP loc' (presented in main text), the  `hierarchy PL loc' and `hierarchy PL' based on a power law deterrence function also outperform the state-of-the-art radiation and gravity PL models.

\begin{table*}[ht!]
\begin{center}
\caption{ {\bf Benchmark goodness fit measures} \label{table:benchmark}
The Dice (D), Sorensen (S), Cosine (C) and deviance (E) are four different fit values measuring a distance between the actual and modeled networks. The correlation $corr$ measures a similarity between the model and the data. The different parameters of the gravity and hierarchy models were chosen to minimize the value of E. Stars (*) denote models where we imposed the population exponents $\alpha$ and $\beta$ to be equal to 1. The rows corresponding to models presented in the main text are highlighted. \\}
\begin{tabular}{l | c | ccc | c | l}
    Country / Model & E$*10^{-9}$ & D & S & C & corr & params\\ 
    \hline \hline
    \rowcolor{myblue} UK / radiation   & 1622.9 &  0.624 & 0.632 & 0.344 &   0.656  &  \\
    UK / gen. radiation   & \cellcolor{white}  818.3 & 0.236 & 0.444 & 0.195 & 0.805 & $1-\lambda = 8.65 \times10^{-10}$\\        
    \hline
    UK / gravity EXP & 941.1 & 0.777 & 0.579 & 0.595  & 0.402 & $\alpha=0.78$, $\beta =0.78$, $d_0=63.8km$\\
    UK / hierarchy EXP & 547.7 &  0.278 & 0.447 & 0.278  & 0.720 & $\alpha=0.90$, $\beta =0.90$, $q=0.138$\\ 
    UK / hierarchy-admin EXP  & 741.2 & 0.580 & 0.536 & 0.512 & 0.485 & $\alpha=0.90$, $\beta=0.90$, $q=0.374$\\
     \arrayrulecolor{gray}\hline\arrayrulecolor{black}
    UK / gravity EXP*  & 969.4 &  0.749 & 0.583 & 0.612 & 0.384 &  $d_0=67.3km$\\
    UK / hierarchy EXP*  & 553.8 & 0.291 & 0.448 & 0.288 & 0.711 & $q=0.139$\\
    UK / hierarchy-admin EXP* & 747.7 & 0.566 & 0.534 & 0.520 & 0.477 &  $q=0.235$\\
    \arrayrulecolor{gray}\hline\arrayrulecolor{black}
    UK / gravity EXP loc  & 839.6  & 0.726 & 0.555 & 0.547 & 0.450 & $d_0=56.1km$\\
    \rowcolor{myblue} UK / hierarchy EXP loc & 464.9 & 0.233 & 0.437 & 0.231 & 0.768 & $q=0.139$\\
    \rowcolor{myblue} UK / hierarchy-admin EXP loc & 662.5 & 0.558 & 0.527 & 0.470 & 0.239 & $q=0.239$\\       
    \hline        
    \rowcolor{myblue} UK / gravity PL & 494.7   &  0.456 & 0.448 & 0.456 & 0.543 & $\alpha=0.65$, $\beta =0.65$, $\gamma=-1.46$\\
    UK / hierarchy PL  & 649.9 &  0.334 & 0.486 & 0.326  & 0.673 & $\alpha=0.91$, $\beta =0.91$, $\gamma=-4.23$ \\
    UK / hierarchy-admin PL & 768.7 & 0.571 & 0.542 & 0.519  & 0.478 & $\alpha=0.90$, $\beta =0.90$, $\gamma=-3.10$ \\
    \arrayrulecolor{gray}\hline\arrayrulecolor{black}
    UK / gravity PL*  & 562.8  & 0.433 & 0.467 & 0.431& 0.567 &  $\gamma=-1.36$\\
    UK / hierarchy PL*  & 655.2 & 0.354 & 0.488 & 0.334 & 0.665 & $\gamma=-4.21$\\
    UK / hierarchy-admin PL* & 774.8 & 0.560 & 0.541 & 0.527 & 0.471& $\gamma=-3.09$\\
    \arrayrulecolor{gray}\hline\arrayrulecolor{black}
    UK / gravity PL loc  & 351.5 & 0.218 & 0.375 & 0.216 & 0.783 & $\gamma=-1.57$\\
    UK / hierarchy PL loc & 555.6 & 0.290 & 0.481 & 0.288  & 0.711 & $\gamma=-4.22$\\
    UK / hierarchy-admin PL loc & 679.9 & 0.540 & 0.533 & 0.468 & 0.529 & $\gamma=-3.08$ \\           
    \hline     \hline
    \rowcolor{myblue} Portugal / radiation   & 314.1 &   0.781 & 0.739 & 0.476 &  0.525 &  \\
    Portugal / gen. radiation   &  \cellcolor{white} 78.4 &  0.472 & 0.420 & 0.423 &  0.563 & $1-\lambda = 4.22 \times10^{-10}$\\    
    \hline
    Portugal / gravity EXP  & 96.20 & 0.718 &  0.461 &  0.577  & 0.400 & $\alpha=0.87$, $\beta =0.86$, $d_0=92.9km$\\
    Portugal / hierarchy EXP & 71.86  & 0.465 & 0.420 & 0.422 & 0.564 & $\alpha=0.91$, $\beta =0.89$, $q=0.244$\\ 
    Portugal / hierarchy-admin EXP & 83.98  & 0.605 & 0.441 & 0.501 & 0.480 & $\alpha=0.83$, $\beta =0.82$, $q=0.343$\\ 
     \arrayrulecolor{gray}\hline\arrayrulecolor{black}
    Portugal / gravity EXP*  & 97.15 & 0.700 & 0.460 & 0.577 & 0.399 &  $d_0=95.6km$\\
    Portugal / hierarchy EXP* & 72.45 & 0.455 & 0.419 & 0.426 & 0.561& $q=0.281$\\
    Portugal / hierarchy-admin EXP*  & 85.65  & 0.580 & 0.441 & 0.500 & 0.482 & $q=0.352$\\
    \arrayrulecolor{gray}\hline\arrayrulecolor{black}
    Portugal / gravity EXP loc  & 92.08  & 0.684 & 0.455 & 0.546 & 0.433 & $d_0=82.0km$\\
    \rowcolor{myblue} Portugal / hierarchy EXP loc & 66.66 &0.346 & 0.404 & 0.308 & 0.683 & $q=0.258$\\
    \rowcolor{myblue} Portugal / hierarchy-admin EXP loc & 74.20 &0.456 & 0.416 & 0.362 & 0.627 & $q=0.278$\\     
    \hline
    \rowcolor{myblue} Portugal / gravity PL  & 79.80 & 0.865 & 0.419 & 0.844  & 0.145 &$\alpha=0.81$, $\beta =0.79$, $\gamma=-0.71$\\
    Portugal / hierarchy PL  & 77.79 & 0.473 & 0.443 & 0.450  & 0.536  & $\alpha=0.91$, $\beta =0.90$, $\gamma=-2.76$ \\
    Portugal / hierarchy-admin PL & 86.87 & 0.593 & 0.452 & 0.511  & 0.470  & $\alpha=0.85$, $\beta =0.83$, $\gamma=-2.28$ \\    
    \arrayrulecolor{gray}\hline\arrayrulecolor{black}
    Portugal / gravity PL*   & 81.88 & 0.793 & 0.420 & 0.781& 0.205 &$\gamma=-0.69$\\
    Portugal / hierarchy PL*   & 78.32  & 0.468 & 0.441 & 0.454 & 0.532 &  $\gamma=-2.73$\\
    Portugal / hierarchy-admin PL*  & 88.23  & 0.572 & 0.450 & 0.509 & 0.473 &  $\gamma=-2.24$\\    
    \arrayrulecolor{gray}\hline\arrayrulecolor{black}
    Portugal / gravity PL loc  &  66.37 & 0.403 & 0.386 & 0.386 & 0.602 & $\gamma = -0.92$\\
    Portugal / hierarchy PL loc & 73.29 &0.373 & 0.433 & 0.352 & 0.637 & $\gamma = -2.89$\\
    Portugal / hierarchy-admin PL loc & 77.05 & 0.429 & 0.431 & 0.363 & 0.625 & $\gamma = -2.72$\\  
    \hline \hline
    \rowcolor{myblue} France / radiation   &  227.758 &0.618 & 0.647 & 0.270 & 0.730 &  \\
    France / gen. radiation   & 109.123 &0.426 & 0.494 & 0.244 & 0.756 &$1-\lambda= 4.40 \times 10^{-9}$\\       
    \hline
    France / gravity EXP & 139.132 & 0.453 & 0.639 & 0.367 & 0.632 &  $\alpha=0.87$, $\beta =0.86$, $d_0=92.0km$\\
    France / hierarchy EXP & 92.284  & 0.443 & 0.518 & 0.330 & 0.670 &  $\alpha=0.89$, $\beta =0.89$, $q=0.142$\\
    France / hierarchy-admin EXP & 101.248  &0.270 & 0.544 & 0.266 & 0.734 &  $\alpha=0.90$, $\beta =0.90$, $q=0.179$ \\
     \arrayrulecolor{gray}\hline\arrayrulecolor{black}
    France / gravity EXP*  &   142.829 &0.370 & 0.637 & 0.370 & 0.630 & $d_0=98.4km$\\
    France / hierarchy EXP* & 94.876 & 0.616 & 0.519 & 0.368 & 0.632 & $q=0.146$ \\
    France / hierarchy-admin EXP* & 103.187 &0.328 & 0.539 & 0.263 & 0.737 & $q=0.183$\\
     \arrayrulecolor{gray}\hline\arrayrulecolor{black}
    France / gravity EXP loc  &  116.792 &0.378 & 0.611 & 0.276 & 0.724 & $d_0=96.7km$ \\
    \rowcolor{myblue} France / hierarchy EXP loc & 73.524 &0.341 & 0.514 & 0.267 & 0.733 & $q=0.158$\\
    \rowcolor{myblue} France / hierarchy-admin EXP loc & 80.686 & 0.212 & 0.529 & 0.207 & 0.793 & $q=0.192$\\    
    \hline
    \rowcolor{myblue} France / gravity PL & 90.905  & 0.267 & 0.524 & 0.185 & 0.815 &  $\alpha=0.69$, $\beta =0.69$, $\gamma=-1.44$\\
    France / hierarchy PL & 103.181 & 0.572 & 0.560 & 0.427 & 0.573 & $\alpha=0.91$, $\beta =0.91$, $\gamma=-4.13$\\
    France / hierarchy-admin PL  & 110.282 & 0.329 & 0.575 & 0.323 & 0.677 & $\alpha=0.91$, $\beta =0.91$, $\gamma=-3.63$ \\
    \arrayrulecolor{gray}\hline\arrayrulecolor{black}
    France / gravity PL*  & 111.425  &0.663 & 0.590 & 0.313 & 0.687 & $\gamma=-1.16$\\
    France / hierarchy PL*  & 105.061  &0.702 & 0.558 & 0.462 & 0.538 & $\gamma=-4.08$\\
    France / hierarchy-admin PL*  & 111.902 & 0.377 & 0.569 & 0.321 & 0.679 & $\gamma=-3.59$\\
    \arrayrulecolor{gray}\hline\arrayrulecolor{black}
    France / gravity PL loc  & 71.101 & 0.237 & 0.495 & 0.123 & 0.877 & $\gamma=-1.36$\\
    France / hierarchy PL loc & 81.804 &0.485 & 0.559 & 0.378  & 0.622 & $\gamma=-3.94$\\
    France / hierarchy-admin PL loc & 87.132 &  0.248 & 0.559 & 0.246 & 0.679 & $\gamma=-3.52$\\        
    \hline 
\end{tabular}
\end{center}
\end{table*}

\begin{table*}[ht!]
\begin{center}
\caption{ {\bf Benchmark goodness fit measures} \label{table:benchmark2}
The Dice (D), Sorensen (S), Cosine (C) and deviance (E) are four different fit values measuring a distance between the actual and modeled networks. The correlation $corr$ measures a similarity between the model and the data. The different parameters of the gravity and hierarchy models were chosen to minimize the value of E. Stars (*) denote models where we imposed the population exponents $\alpha$ and $\beta$ to be equal to 1.\\}
\begin{tabular}{l | c | ccc | c | l}
    Country / Model &  E$*10^{-9}$ & D & S & C & corr & params\\ 
    \hline \hline
     \rowcolor{myblue} Country X / radiation  & 3.701 &  0.577 & 0.638 & 0.356  & 0.644 & \\
    Country X / gen. radiation   &  1.396 &  0.244 & 0.415 & 0.241 & 0.759 &$1-\lambda = 4.49\times 10^{-7}$\\
    \hline
    Country X / gravity EXP & 2.167 & 0.760 & 0.531 & 0.558  & 0.444 & $\alpha=0.83$, $\beta =0.80$, $d_0=53.2km$\\
    Country X / hierarchy EXP & 1.508 & 0.361 & 0.483 & 0.348  & 0.651 & $\alpha=0.90$, $\beta =0.87$, $q=0.117$\\ 
    Country X / hierarchy-admin EXP & 2.439 & 0.817 & 0.585 & 0.645  & 0.354 & $\alpha=0.82$, $\beta = 0.79$, $q=0.156$\\ 
     \arrayrulecolor{gray}\hline\arrayrulecolor{black}
    Country X / gravity EXP*  & 2.333 & 0.698 & 0.533& 0.555 & 0.442 &  $d_0=54.3km$\\
    Country X / hierarchy EXP* & 1.534 & 0.340 & 0.483 & 0.340 & 0.660&  $q=0.118$\\
    Country X / hierarchy-admin EXP* & 2.512 & 0.765 & 0.587 & 0.642 & 0.356 & $q=0.159$\\ 
    \arrayrulecolor{gray}\hline\arrayrulecolor{black}
    Country X / gravity EXP loc  & 1.915  & 0.665 & 0.488 & 0.478 & 0.522 & $d_0=50.2km$\\
    \rowcolor{myblue} Country X / hierarchy EXP loc & 1.120 & 0.255 & 0.456 & 0.252 & 0.748 & $q=0.114$\\
    \rowcolor{myblue} Country X / hierarchy-admin EXP loc & 2.076 & 0.743 & 0.547 & 0.565 & 0.434 & $q=0.158$\\  
    \hline
    \rowcolor{myblue} Country X / gravity PL  & 1.483 & 0.472 & 0.467 & 0.470 &   0.529  & $\alpha=0.81$, $\beta =0.78$, $\gamma=-1.06$\\
    Country X / hierarchy PL  & 2.021 & 0.389 & 0.565 & 0.384   & 0.615 & $\alpha=0.91$, $\beta =0.88$, $\gamma=-4.43$ \\
    Country X / hierarchy-admin PL & 2.257 & 0.794 & 0.567 & 0.638  & 0.361  & $\alpha=0.83$, $\beta =0.80$, $\gamma=-3.71$\\ 
    \arrayrulecolor{gray}\hline\arrayrulecolor{black}
    Country X / gravity PL* & 1.573 & 0.409 & 0.469 & 0.405 & 0.595 & $\gamma=-1.07$\\
    Country X / hierarchy PL*  & 2.042 & 0.378 & 0.566 & 0.378 & 0.622 &  $\gamma=-4.41$\\
    Country X / hierarchy-admin PL* & 2.322 & 0.742 & 0.567& 0.636 & 0.363 & $\gamma=-3.67$\\ 
    \arrayrulecolor{gray}\hline\arrayrulecolor{black}
    Country X / gravity PL loc  & 0.940 & 0.202 & 0.382 & 0.201 & 0.798 & $\gamma=-1.22$\\
    Country X / hierarchy PL loc & 1.589 & 0.310 & 0.556 & 0.308 &0.692 & $\gamma=-4.43$\\
    Country X / hierarchy-admin PL loc & 1.882 & 0.707 & 0.523 & 0.545 & 0.454 & $\gamma=-3.77$\\  
    \hline \hline
    \rowcolor{myblue} Ivory Coast / radiation   & 268.18 & 0.701 & 0.703 & 0.358 & 0.645 & \\
    Ivory Coast / gen. radiation   &  \cellcolor{white}  72.08 &0.373 & 0.442 & 0.349 & 0.638 & $1-\lambda = 4.20 \times10^{-10}$\\       
    \hline
    Ivory Coast / gravity EXP & 74.02 & 0.680 & 0.430 & 0.539 & 0.434 & $\alpha=0.96$, $\beta =0.96$, $d_0=149.6km$ \\
    Ivory Coast / hierarchy EXP & 45.85 & 0.297 & 0.366 & 0.374 & 0.716 & $\alpha=0.93$, $\beta =0.93$, $q=0.271$ \\
    Ivory Coast / hierarchy-admin EXP & 78.71 & 0.685 & 0.448 & 0.553 & 0.418 & $\alpha=0.96$, $\beta =0.96$, $q=0.567$\\
     \arrayrulecolor{gray}\hline\arrayrulecolor{black}
    Ivory Coast / gravity EXP*  & 74.08 & 0.673 & 0.429 & 0.537 & 0.436 & $d_0=149.7km$ \\
    Ivory Coast / hierarchy EXP* & 46.15 & 0.285 & 0.364 & 0.270 & 0.720 &  $q=0.273$ \\
    Ivory Coast / hierarchy-admin EXP* & 78.78 & 0.678 & 0.447 & 0.550 & 0.421 &  $q=0.568$\\
    \arrayrulecolor{gray}\hline\arrayrulecolor{black}
    Ivory Coast / gravity EXP loc  & 65.51 & 0.567 & 0.414 & 0.423 & 0.561 & $d_0=112.9km$ \\
    \rowcolor{myblue}  Ivory Coast / hierarchy EXP loc & 42.90  & 0.228 & 0.351 & 0.217 & 0.775 & $q=0.255$ \\
    \rowcolor{myblue} Ivory Coast / hierarchy-admin EXP loc & 65.98 &0.437 & 0.430 & 0.309 & 0.681 & $q=0.394$ \\  
    \hline
    \rowcolor{myblue}  Ivory Coast / gravity PL  & 68.17 & 0.577 & 0.413 & 0.460 & 0.519 & $\alpha=0.94$, $\beta =0.94$, $\gamma =-0.51$ \\
    Ivory Coast / hierarchy PL  & 50.89 & 0.321 & 0.378 & 0.317 & 0.672 & $\alpha=0.93$, $\beta =0.93$, $\gamma =-2.89$ \\
    Ivory Coast / hierarchy-admin PL  & 80.72 & 0.692 & 0.450 & 0.558 & 0.412 & $\alpha=0.96$, $\beta =0.96$, $\gamma =-1.19$\\
    \arrayrulecolor{gray}\hline\arrayrulecolor{black}
    Ivory Coast / gravity PL*  & 68.37 & 0.561 & 0.411 & 0.453 & 0.527 &  $\gamma=-0.51$  \\
    Ivory Coast / hierarchy PL*  & 51.18 & 0.315 & 0.376 & 0.313 & 0.676 & $\gamma =-2.88$\\
    Ivory Coast / hierarchy-admin PL*  & 80.80 & 0.684 & 0.449 & 0.555 & 0.415 & $\gamma =-1.19$\\
    \arrayrulecolor{gray}\hline\arrayrulecolor{black}
    Ivory Coast / gravity PL loc  & 54.03  & 0.267 & 0.374 & 0.240 & 0.752 & $\gamma = -0.76$\\
    Ivory Coast / hierarchy PL loc & 48.69 & 0.276 & 0.371 & 0.275 & 0.716 & $\gamma = -2.99$\\
    Ivory Coast / hierarchy-admin PL loc & 68.53 &0.445 & 0.438 & 0.317 & 0.674 & $\gamma = -1.99$\\  
    \hline
\end{tabular}
\end{center}
\end{table*}

\begin{table*}[t!]
\begin{center}
\caption{{\bf Models ranked} according to their average performance across all studied countries, for each benchmark measure (the deviance $E$ being normalized by its value for the Hierarchy EXP loc model $E_{ref}$). Models are sorted according to $\langle rank_{E/E_{ref}}\rangle$ (note: this ranking does not take into account the number of parameters involved in the different models). The rows corresponding to models presented in the main text are highlighted. \label{tab:rankModel}}
\begin{tabular}{l | rrrrr | r }
Model  & $\langle rank_{E/E_{ref}} \rangle$ &  $\langle rank_{D} \rangle$  & $\langle rank_{S} \rangle$ & $\langle rank_{C} \rangle$  & $\langle rank_{corr} \rangle$  & $\langle rank \rangle$ \\
\hline
\rowcolor{myblue} hierarchy EXP loc & 1.80 & 3.00 & 2.20 & 3.20 & 3.80 & 2.80\\
gravity PL loc & 2.20 & 1.80 & 2.00 & 2.20 & 3.00 & 2.24\\
hierarchy EXP & 4.20 & 7.00 & 4.40 & 8.00 & 7.40 & 6.20\\
hierarchy EXP* & 5.20 & 7.40 & 4.40 & 7.40 & 7.80 & 6.44\\
hierarchy PL loc & 5.40 & 5.80 & 8.20 & 6.60 & 7.00 & 6.60\\
\rowcolor{myblue} gravity PL & 6.80 & 11.40 & 5.60 & 11.00 & 12.00 & 9.36\\
\rowcolor{myblue} hierarchy-admin EXP loc & 8.20 & 8.80 & 8.60 & 7.80 & 10.00 & 8.68\\
hierarchy PL & 8.40 & 9.60 & 9.80 & 10.00 & 10.60 & 9.68\\
hierarchy-admin PL loc & 8.80 & 8.40 & 9.80 & 8.00 & 10.00 & 9.00\\
hierarchy PL* & 9.60 & 9.80 & 9.40 & 10.20 & 10.60 & 9.92\\
gravity PL* & 10.40 & 13.40 & 7.80 & 12.00 & 12.80 & 11.28\\
gen. radiation & 11.00 & 6.80 & 5.00 & 4.40 & 5.60 & 6.56\\
gravity EXP loc & 11.60 & 13.60 & 12.20 & 13.20 & 13.60 & 12.84\\
hierarchy-admin EXP & 13.60 & 14.60 & 12.40 & 14.00 & 14.40 & 13.80\\
hierarchy-admin EXP* & 14.80 & 13.00 & 12.00 & 13.60 & 14.00 & 13.48\\
hierarchy-admin PL & 15.20 & 14.60 & 14.60 & 15.80 & 16.20 & 15.28\\
hierarchy-admin PL* & 16.40 & 13.20 & 13.80 & 15.40 & 15.80 & 14.92\\
gravity EXP & 16.20 & 16.80 & 14.00 & 16.00 & 16.20 & 15.84\\
gravity EXP* & 17.60 & 14.20 & 13.80 & 16.20 & 16.80 & 15.72\\
\rowcolor{myblue} radiation & 19.80 & 16.80 & 18.00 & 8.80 & 9.40 & 14.56\\
\end{tabular}
\end{center}
\end{table*}
\clearpage 
\bibliographystyle{plain}

\bibliography{refs}